\documentclass[aps, pre, twocolumn, superscriptaddress]{revtex4-1}

\usepackage{amssymb, amsmath, amsfonts, amsthm, mathtools}
\usepackage{graphicx, xcolor}
\usepackage{empheq}
\usepackage[T1]{fontenc}
\usepackage{xspace}
\usepackage{epstopdf}

\usepackage{soul}

\renewcommand*{\d}{{\mathrm d}}

\newcommand*{\xth}{x_\mathrm{th}\xspace}

\newcommand{\eq}[1]{Eq.~#1}
\newcommand{\eqs}[1]{Eqs.~#1}
\newcommand{\fig}[1]{Fig.~#1}
\newcommand{\figs}[1]{Figs.~#1}

\begin{document}

\title{Evolution of moments and correlations in non-renewal escape-time processes}

\author{Wilhelm Braun}
\email{wilhelm.braun@cantab.net}
\affiliation{Department of Physics and Centre for Neural Dynamics, University of Ottawa, 598 King Edward, Ottawa K1N 6N5, Canada}
\author{R\"{u}diger Thul}
\affiliation{Centre for Mathematical Medicine and Biology, School of Mathematical Sciences, University of Nottingham, Nottingham, NG7 2RD, UK}
\author{Andr\'{e} Longtin}
\affiliation{Department of Physics and Centre for Neural Dynamics, University of Ottawa, 598 King Edward, Ottawa K1N 6N5, Canada}

\date{\today}

\begin{abstract}  
The theoretical description of non-renewal stochastic systems is a challenge. Analytical results are often not available or can only be obtained under strong conditions, limiting their applicability. Also, numerical results have mostly been obtained by ad-hoc Monte--Carlo simulations, which are usually computationally expensive when a high degree of accuracy is needed. To gain quantitative insight into these systems under general conditions, we here introduce a numerical iterated first-passage time approach based on solving the time-dependent Fokker--Planck equation (FPE) to describe the statistics of non-renewal stochastic systems.  We illustrate the approach using spike-triggered neuronal adaptation in the leaky and perfect integrate-and-fire model, respectively. The transition to stationarity of first-passage time moments and their sequential correlations occur on a non-trivial timescale that depends on all system parameters. Surprisingly this is so for both single exponential and scale-free power-law adaptation. The method works beyond the small noise and timescale separation approximations. It shows excellent agreement with direct Monte Carlo simulations, which allows for the computation of transient and stationary distributions. We compare different methods to compute the evolution of the moments and serial correlation coefficients (SCC), and discuss the challenge of reliably computing the SCC which we find to be very sensitive to numerical inaccuracies for both the leaky and perfect integrate-and-fire models. In conclusion, our methods provide a general picture of non-renewal dynamics in a wide range of stochastic systems exhibiting short and long-range correlations. 
\end{abstract}


\maketitle

\section{Introduction}

A general property of diverse systems, ranging from superconducting quantum interference devices (SQUIDs) \cite{palacios_et_al_2006}, to lasers \cite{aragoneses_et_al_2016} to excitable cells \cite{chacron_lindner_longtin_2004, reinoso_torrent_masoller_2016, coombes_thul_noisy_threshold_2011, nesse_maler_longtin_2010} is that time intervals between specific events are not statistically  independent. The theoretical description of such non-renewal stochastic processes \cite{cox_lewis_1966} poses a significant challenge, as it implies that the present state of the system depends, in general, on the whole past evolution or parts of it, and not just on the previous state. Analytical approximations to tackle such memory effects have included the assumption of stationarity \cite{rosenbaum_2016}, small stochasticity \cite{schwalger_linder_pre_2015}  and time-scale separation \cite{liu_wang_2001, muller_et_al_2007, hertaeg_durstewitz_brunel_2014} between stochastic and deterministic parts of the dynamics.

Even if these approximations allow for some insight into the parameter dependence of e.g. serial correlations and can be used to understand experimental data, as exemplified in \cite{schwalger_linder_pre_2015, Schwalger_et_al_2010, fisch_et_al_2012, urdapilleta_pre_onset_of_correlations_2011, urdapilleta_2016} in the context of excitable systems, it is desirable to understand the statistics of model systems without making simplifying assumptions. Regarding stationarity, real systems rarely operate in a stationary state due to transients that arise from deterministic or random perturbations. A prominent example are cortical neurons. An average cortical neuron receives random inputs from approximately $10^{4}$ other neurons, whose activity is modulated by non-stationary sensory and other inputs, resulting in transient neuronal dynamics \cite{naud_gerstner_2012} that only become stationary after a certain time. It is therefore important to understand how statistical properties of inter-event times evolve and become invariant following a transient regime due to internal dynamics and external inputs. Keeping with illustrations from neural dynamics, it is well known that physiologically relevant processes underlying neural coding rarely only have one well-defined timescale \cite{la_camera_et_al_2006, ulanovsky_et_al_2004}. This has lead researchers in various theoretical fields to consider multiple time-scale dynamics \cite{CKbook}. An important example of a system with multiple time scales is neuronal adaptation, where a neuron's firing rate adjusts in response to a stimulus. Adaptation with multiple timescales, or even no time scale as in the case of power-law adaptation \cite{drew_abbott_2006, lundstrom_et_al_2008, pozzorini_naud_mensi_gerstner_2013}, is now known to be biophysically relevant, and even optimal for some tasks \cite{clarke_pnas_2013}. Recently, it was also shown that a neuron model with adaptive firing thresholds exhibiting multiple timescales is the optimal choice for the prediction of spike times in cortical neurons \cite{kobayashi_2009, gerstner_naud_2009}. Therefore, a theoretical description of adaptation without a single well-defined timescale is an important goal.

In this paper, we show how to describe two-dimensional non-renewal dynamics by an iterated first-passage time (iFPT) approach. This approach allows us to determine stationary statistical properties of the system as well as providing a description of the transition to stationarity. We furthermore show how to compute serial correlations in the time series generated by the firing times of the system. While our approach is general and applicable to any system where first-passage times \cite{redner_book} play a role, we illustrate its versatility with two important examples, namely spike-triggered neuronal adaptation with a single exponential current and a power-law current without an intrinsic timescale, respectively. Using the underlying time-dependent FPE to describe the system, we only need to apply mathematically convenient standard absorbing boundary conditions to obtain stationary distributions, e.g. that of the adaptation current upon firing. Moreover, the methods developed here can easily be extended to models with correlation-generating deterministic input currents as recently considered in \cite{donofrio_pirozzi_magnasco_2016}.

\section{Model}

We consider a stochastic differential equation (SDE) driven by an external signal $s(t)$:
\begin{equation}
\d X(t) = \mu(X(t))d t + \phi(X(t))\d W(t) - s(t) \d t \, .
\label{eq:X_definition}
\end{equation}

$X$ is defined on the domain $(-\infty, \xth]$. If $X$ reaches $\xth$, the system is said to have generated and event, and $X$ is instantaneously reset to $0$. For all examples in this study, we chose the Ornstein--Uhlenbeck process (OUP) given its prominence in the field of stochastic systems. For the OUP, we fix the correlation time $\tau_{m} = \frac{1}{\gamma}$, bias current $I_{0}$ and noise intensity $\sigma$ as follows: $\mu(X(t)) = \gamma(I_{0} - X(t))$, $\phi(X(t)) =\sigma \gamma$. $W(t)$ is a standard Brownian motion and we set $\xth = 1$. Given that the OUP is the basis for integrate-and-fire (IF) neuron models, which are among the most popular neuron descriptions \cite{burkitt_review_1}, we refer to events as spikes and to $s(t)$ as a time-dependent adaptation current in the present study. The general dynamics of $s(t)$ obeys a single autonomous ordinary differential equation (ODE)

\begin{equation}
\dot{s} = \omega(s)\, ,
\label{eq:s_definition}
\end{equation}

and $s$ is increased by a fixed amount $\kappa$ when $X  = \xth$: $s \rightarrow s + \kappa$, which is the mechanism for spike-triggered adaptation \cite{benda_herz_2003}. When $s(t)$ is also reset to its starting value $s(0)$, the model is a renewal model and its firing statistics may be studied using standard techniques, see e.g. \cite{gs_review} for a recent review. 

Here we focus on two forms of the adaptation current. The first one is power-law adaptation, for which 

\begin{equation}
\omega(s) = -\frac{1}{\alpha} s^{2}(t) \, .
\label{eq:power_law}
\end{equation}

This ODE has the general solution $s(t) = \left( \frac{t}{\alpha} + \frac{1}{s(0)}\right)^{-1}$. Therefore, the current $s$ in this case has a power-law time dependence with no intrinsic time scale \cite{drew_abbott_2006}. 

The second adaptation current is given by a single exponential decay with time scale $\tau_{a}$:

\begin{equation}
\omega(s) = -\frac{1}{\tau_{a}} s(t) \, ,
\label{eq:single_exponential}
\end{equation}

which has the general solution $s(t) = s(0)e^{-\frac{t}{\tau_{a}}}$.

The time to the first spike event is the following first-passage time (FPT):

\begin{equation*}
T_{1} = \inf(t>0: X(t) > \xth | X(0) =0, s(t=0) = s(0)) \, .
\end{equation*}

\begin{figure}[h!]
\centering
\includegraphics[width = 0.5\textwidth]{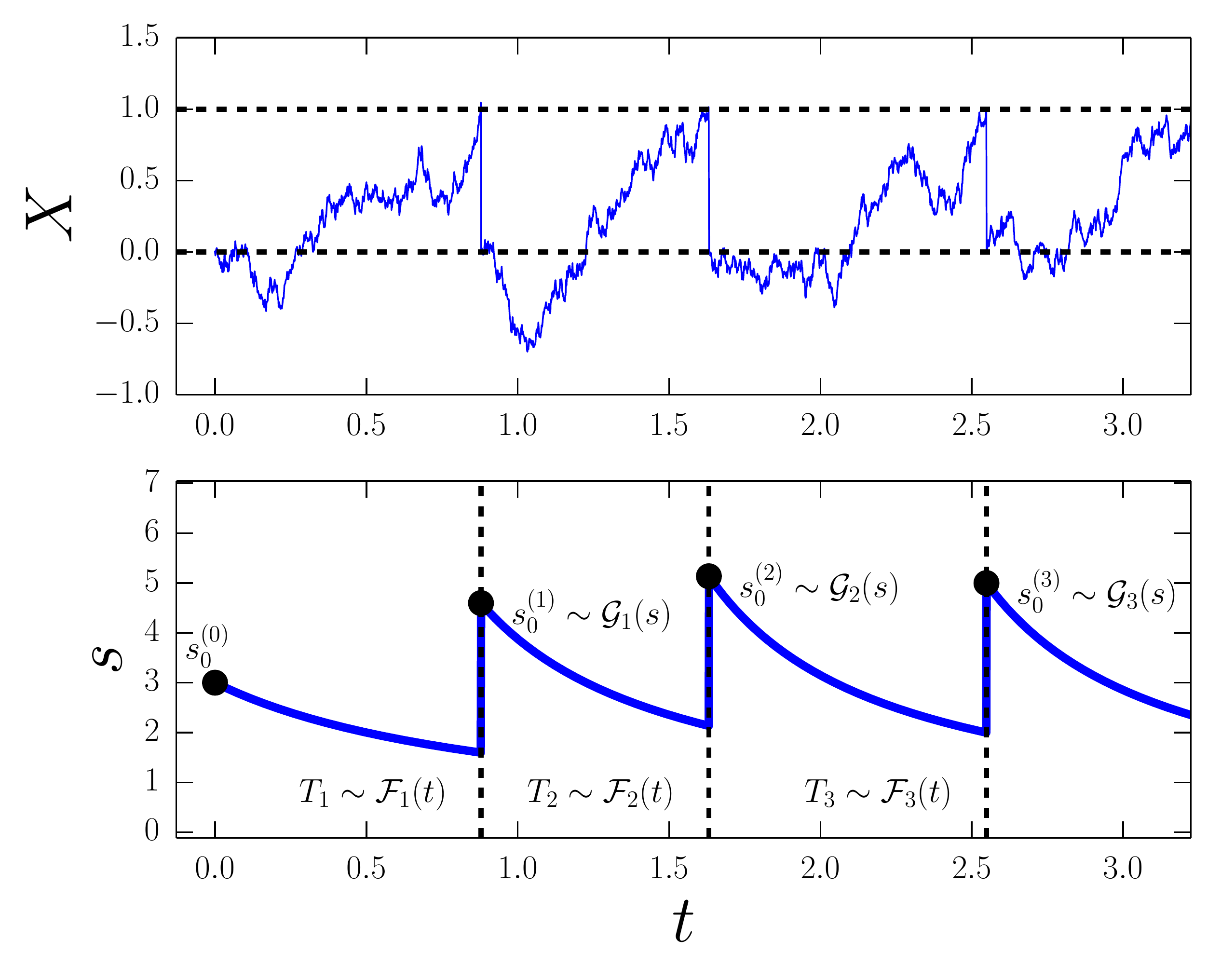}
\caption{Sample paths of the model for a power-law adaptation current given by \eq \ref{eq:power_law}. \textbf{Top:} $X(t)$ (\eq 
\ref{eq:X_definition}), the horizontal dashed lines are at $X = 0$ and $\xth= 1$. \textbf{Bottom}: $s(t)$ (\eq \ref{eq:s_definition}) When $X$ reaches $\xth$, $s$ undergoes a jump of size $\kappa$. The subsequent ISIs $T_{k}$ have distributions $\mathcal F_{k}(t)$, and the starting values $s^{(k)}_{0}$ have distributions $\mathcal G_{k}(s)$ for $k \geq 1$. Parameter values are $\alpha = 3.0,~\gamma = 1.0,~\sigma = 0.8 ,~I_{0} = 4.0,~\kappa = 3.0$.}
\label{fig:ensemble_sketch}
\end{figure}

In the non-renewal case we are studying here, subsequent firing times will in general not have the same distribution as $T_{1}$. We define the $k$th interspike interval (ISI) as
\begin{equation}
T_{k} = \inf\left(t- \sum_{i=1}^{k-1}T_{i} ~:~ X(t) \geq \xth, t > \sum_{i=1}^{k-1}T_{i}\right) \, .
\label{eq:kth_firing_time}
\end{equation}
The first moment of the $k$th ISI is given by $\tau^{1}_{k}= \mathbb E(T_{k})$. The second moment of the $k$th ISI will be denoted by $\tau^{2}_{k}= \mathbb E((T_{k})^{2})$ and the $k$th firing rate is given by the inverse of the corresponding mean ISI $r_{k} = \frac{1}{\tau^{1}_{k}}$. The $k$th standard deviation $m_{2}(k)$ is then given by

\begin{equation}
m_{2}(k) = \sqrt{\tau^{2}_{k}-(\tau^{1}_{k})^{2}} \, .
\label{eq:m2_definition}
\end{equation}

The values of the peak adaptation current after the $k$th firing are defined for $k\geq 1$ as

\begin{equation}
s_{0}^{(k)} = \left( s(t^{-}) + \kappa ~:~ t = \sum_{i=1}^{k}T_{i} \right) \, ,
\label{eq:kth_adaptation_current}
\end{equation}

where $t^{-}$ indicates that we take the left-sided limit.
For simplicity, we choose $s$ to be started from a point ($s_{0}^{(0)} = \kappa$), instead of from a biophysically more realistic initial distribution. However, the methods we are going to describe in the following are also valid when $s$ is initially started from a distribution.

The central challenge is to obtain the distributions $\mathcal G_{k}$ and $\mathcal F_{k}$ for $k \geq 1$, which are the distributions of $s_{0}^{(k)}$ and $T_{k}$ defined by \eqs \ref{eq:kth_adaptation_current} and \ref{eq:kth_firing_time}, respectively. An example realization for the case of power-law adaptation is shown in \fig \ref{fig:ensemble_sketch}. The knowledge of these distributions is key to understanding the non-renewal dynamics, as they form a hidden Markov model of the underlying non-Markovian dynamics  \cite{van_vreeswijk_2010, urdapilleta_pre_onset_of_correlations_2011, urdapilleta_2016}. Therefore, once these distributions are known, the non-renewal dynamical system breaks up into coupled renewal dynamical systems, which are much more tractable mathematically. This gives rise to the iFPT approach which we now explain.

\section{The iFPT approach}
Being a diffusion process, the system given by \eqs \ref{eq:X_definition} and \ref{eq:s_definition} is governed by a two-dimensional time-dependent FPE \cite{risken_book}. The FPE for the probability density function $p(t;x,s)\d x \d s = \mathbb P(X(t) \in (x, x+ \d x), s(t) \in (s, s + \d s) | X(0) = x_{0}, s(0) = s_{0})$ reads (we use $\mathbf{x}^{\top} = (x,s)$ for brevity)

\begin{equation}
\partial_{t} p(t; \mathbf{x}) = \nabla \cdot (\mathbf{A(\mathbf{x})} \nabla p(t; \mathbf{x})) -\nabla \cdot \left(\mathbf{F(\mathbf{x})} p\right(t; \mathbf{x})) \, ,
\label{eq:FPE_general} 
\end{equation}

where $\mathbf{A}$ is the diffusion matrix and $\mathbf{F}$ the drift vector, which can be obtained in a straightforward way from the SDE for $X$, \eq \ref{eq:X_definition},  and the ODE for $s$, \eq \ref{eq:s_definition}. Explicitly, we have

\begin{equation}
 \mathbf{A}(\mathbf{x})= 
 \begin{pmatrix}  
  \frac{\phi(x)^{2}}{2} & 0 \\
  0  & 0 
 \end{pmatrix} \, ,
\label{eq:diffusion_matrix}
\end{equation} 

and 

\begin{equation*}
 \mathbf F(\mathbf{x}) = \left(\mu(x) -  s,  \omega(s) \right)^{\top} \, .
\end{equation*}

The IF property of $X$ entails that we have an absorbing boundary at $X = \xth$ for all times $t$: $p(t; x = \xth, s) =0$. With this boundary condition, we can compute the cumulative distribution function (CDF) of the first-passage time $T_{1}$ $, \mathrm{CDF}_{1}(t) = \int_{0}^{t} \mathcal F_{1}(\lambda) \d \lambda $, by time evolution of the FPE on a computational domain $\Omega \subset \mathbb R^{2}$, which we choose to be a rectangle extending to sufficiently negative values in the $x$-direction \cite{spencer_bergman_1993}:

\begin{equation}
 \mathrm{CDF}_{1}(t) = 1 - \int_{\Omega} p_{1}(t;\mathbf{z}) \d \mathbf{z} \, ,
 \label{eq:CDF_numerical_definition}
\end{equation}

where $p_{1}$ is the solution to \eq \ref{eq:FPE_general} with the initial condition $p_{1}(0;x,s) = \delta(\mathbf{x} - \mathbf{x}_{0})$, where $\mathbf{x}_{0} = (0, \kappa)^{\top}$. For the computation of $\mathcal F_{1}$, we then only need to differentiate \eq \ref{eq:CDF_numerical_definition}.

To describe adaptation, one needs to compute the statistics of the peak adaptation currents, as defined by \eq \ref{eq:kth_adaptation_current}. Hence, we need to characterize the hidden Markov model generated by the ISIs $T_{k}$ and the peak adaptation currents $s^{(k)}_{0}$. Whereas the dynamics of $s$ and $X$ as a whole is non-Markovian, the distribution of the ISI $T_{k}$ is completely determined by the distribution $\mathcal G_{k-1}$, and the sequence $\{T_{k}, s_{0}^{(k-1)}\}_{k=1}^{\infty}$ is therefore Markovian. Knowing the values of $T_{k}$ and $s_{0}^{(k-1)}$, the value of $s_{0}^{(k)}$ is fixed, and the distribution of the ISI $T_{k+1}$ can be obtained by solving an FPT problem with $s_{0}^{(k)}$ as initial condition for $s$.

The central observation now is that for the second ISI, $X$ again starts from $0$, whereas $s$ starts from a distribution $\mathcal G_{1}$. This is because $X$ evolves stochastically, and therefore reaches the threshold $\xth$ at different times $T_{1}$, corresponding to different values of $s(t= T_{1}) + \kappa$ immediately after the firing event. To compute the second ISI, we therefore need to know $\mathcal G_{1}$. This can be iterated: to compute the distribution $\mathcal F_{k}$ of the $k$th ISI, we need the distribution $\mathcal G_{k-1}$. This is the central idea of the iFPT approach. To set up the iFPT approach, we first observe that between threshold crossings of $X$, $s$ evolves deterministically. Therefore, when we know the PDF of the first FPT, by conservation of probability, we also know the distribution of $s$ after the first firing event:

\begin{equation}
 \mathcal G_{1}\left(s-\kappa\right) = \left|\frac{\d t(s)}{\d s} \right|\mathcal F_{1}(t(s)) \, ,
\label{eq:G_1}
\end{equation}

where $t(s)$ is the inverse function of $s$. The support of $\mathcal G_{1}$ is shifted, because of the jump of size $\kappa$ that $s$ undergoes when $X$ reaches its threshold $\xth$. 
For the second ISI $T_{2}$, $s$ is started from the distribution $\mathcal G_{1}(s)$ instead of a point, whereas $X$ is started from a point again (\fig \ref{fig:ensemble_sketch}). This means that to obtain $\mathcal F_{2}$, the FPE is started from a distribution: $p_{2}(0;\mathbf{x}) \propto \mathcal G_{1}(s)\delta(x)$. This generalizes to values of $k$ larger than $1$. For the $k$th ISI distribution $\mathcal F_{k}(t)$, we therefore must choose 

\begin{equation}
p_{k}(0;\mathbf{x}) \propto \mathcal G_{k-1}(s)\delta(x) \, . 
\label{eq:p_k_initial_condition}
\end{equation}

We show how to obtain the distributions $\mathcal G_{k}$ for $k>1$ in Section \ref{sec:gk_computation}. Linear splines are used to create a mesh function approximating \eq \ref{eq:p_k_initial_condition} on the computational domain $\Omega$. Due to this approximation, \eq \ref{eq:p_k_initial_condition} then has to be normalized appropriately, so that  $\int_{\Omega} p_{k}(0;\mathbf{z}) \d \mathbf{z} = 1$. The FPE is then solved again, and the distributions $\mathcal F_{k}(t)$ are obtained analogously  to \eq \ref{eq:CDF_numerical_definition}: $\mathrm{CDF}_{k}(t) = 1 - \int_{\Omega} p_{k}(t;\mathbf{z}) \d \mathbf{z}$, i.e. by timestepping the FPE to obtain the CDF of the $k$th ISI followed by a numerical differentiation. This constitutes the iFPT approach.

To quantify the accuracy of our numerical methods, we also compute the relative disagreement $\Delta$ between results obtained by the iFPT approach and direct MC simulations. It is defined for a quantity $Z$ by

\begin{equation}
\Delta(Z) = \frac{\left|Z_{\text{iFPT}}- Z_{\text{MC}} \right|}{Z_{\text{iFPT}}} \, .
\label{eq:relative_error}
\end{equation}

We performed MC simulations for two different simulation setups, the first one without any boundary correction (plain MC), and the second one with a boundary correction according to Giraudo and Sacerdote (MC-GS) \cite{gs_algorithm, gs_review, lindner_longtin_2005}. This boundary correction is applied to reduce the systematic overestimation of FPTs when using the Euler--Maruyama scheme. We compute the relative disagreement given by \eq \ref{eq:relative_error} using either MC simulations with or without boundary correction; we observed that the order of the relative disagreement is unchanged, but in general, the plain MC algorithm gives rise to larger disagreements than MC-GS. A decrease in the relative disagreement is expected, because the GS correction method should yield an improved weak error of $\mathcal O(h)$ \cite{gobet_2000}, in contrast to the plain MC simulation, which has a weak error of $\mathcal O(h^{\frac{1}{2}})$ \cite{higham_mlmc_mfpt}, where $h$ is the time step for the discretization of the SDE, \eq \ref{eq:X_definition}. In the following, the timestep for MC simulations is chosen to be $h = 10^{-3}$ and we choose $M = 10^{6}$ independent realizations. The plain Euler-Maruyama scheme then gives rise to a weak error of $\mathcal O(h^{\frac{1}{2}}) \approx 3 \cdot 10^{-2}$ when estimating moments of first passage times, which is one order of magnitude larger than the MC error proportional to $\frac{1}{\sqrt{M}} = 10^{-3}$. Therefore, in our simulations, the plain MC error is negligible in comparison to the error introduced by the finite-time discretization of the SDE (\eq \ref{eq:X_definition}).

\begin{figure}[h!]
\centering
\includegraphics[width = 0.5\textwidth]{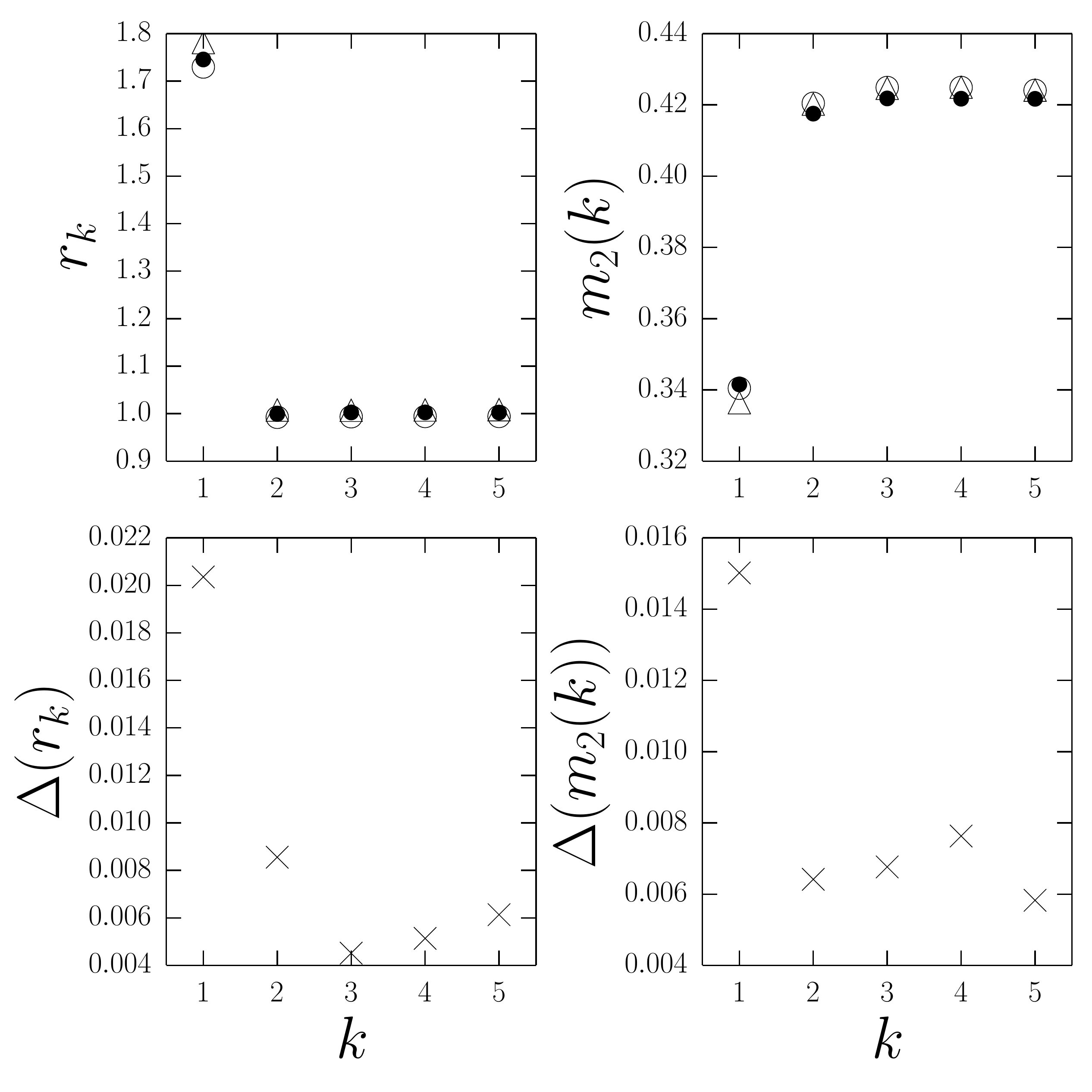}
\caption{\textbf{Top:} Evolution of the rate $r_{k} = \frac{1}{\tau^{1}_{k}}$ (left) and standard deviation given by \eq \ref{eq:m2_definition} (right) of ISIs as a function of $k$. Empty circles: Plain MC simulations of \eqs \ref{eq:X_definition} and \ref{eq:s_definition}. Triangles: MC simulations with GS boundary correction. Filled circles: moments obtained from numerical solution of FPE using a CN timestepping scheme. $M = 10^{6}$ independent MC realizations for each value of $k$. Power-law adaptation (\eq \ref{eq:power_law}) with $\alpha= 5.5,~I_{0} = 6.0$, $\sigma = 1.3$, $\gamma = 1.0, \kappa = 5.5$.  \textbf{Bottom:} Relative disagreements defined by \eq \ref{eq:relative_error}, where an MC-GS algorithm was used.}
\label{fig:moments_evolution_power_law}
\end{figure}

\begin{figure}[h!]
\centering
\includegraphics[width = 0.5\textwidth]{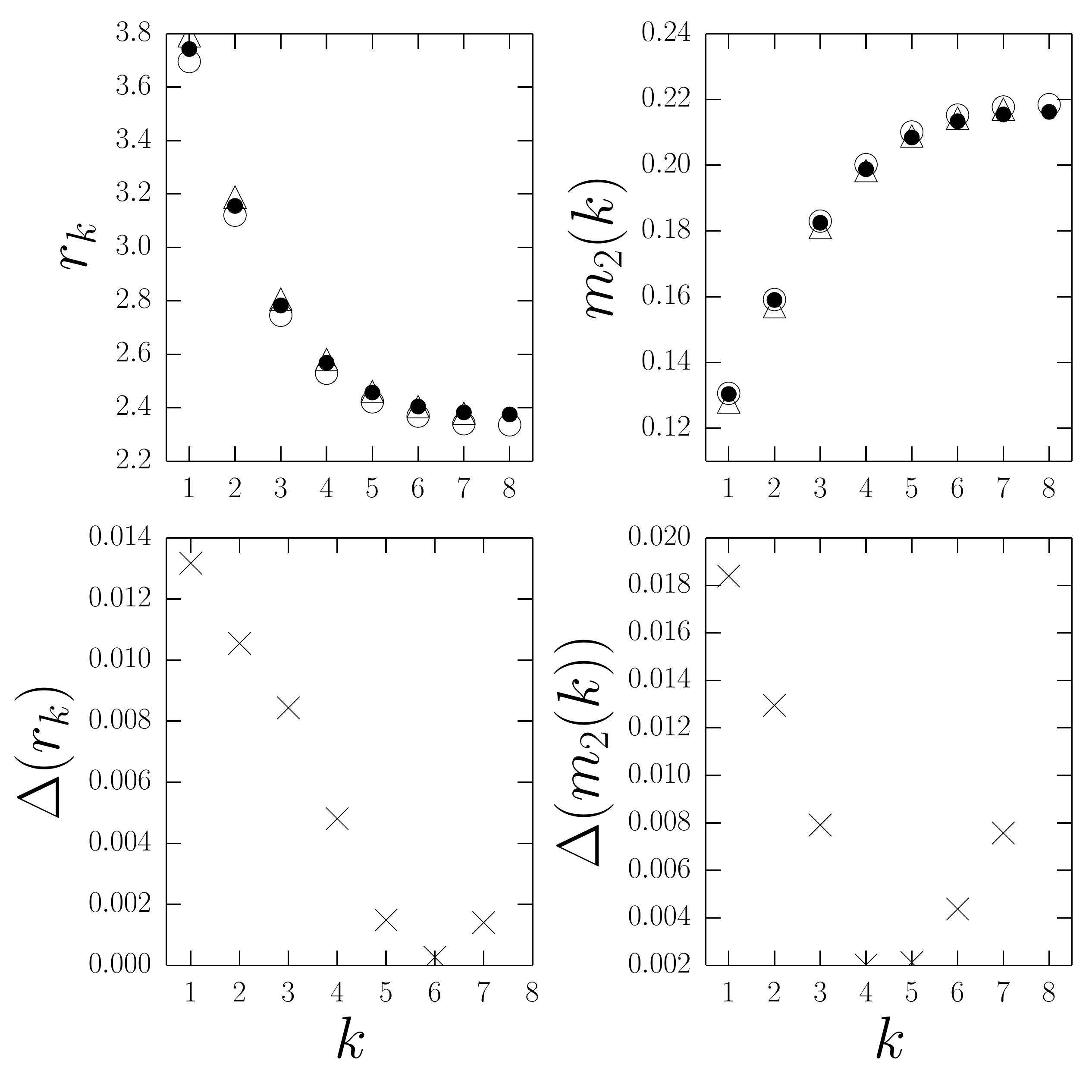}
\caption{\textbf{Top:} Evolution of the rate $r_{k} = \frac{1}{\tau^{1}_{k}}$ (left) and standard deviation given by \eq \ref{eq:m2_definition} (right) of ISIs as a function of $k$. Empty circles: Plain MC simulations of \eqs \ref{eq:X_definition} and \ref{eq:s_definition}. Triangles: MC simulations with GS boundary correction. Filled circles: moments obtained from numerical solution of FPE using an Euler timestepping scheme. $M = 10^{6}$ independent MC realizations for each value of $k$. Single exponential adaptation (\eq \ref{eq:single_exponential}) with $\tau_{a} = 1.0$, $I_{0} = 5.0$, $\sigma = 1.0$, $\gamma = 1.0, \kappa = 1.0$. \textbf{Bottom:}  Relative disagreements defined by \eq \ref{eq:relative_error}, where an MC-GS algorithm was used.}
\label{fig:moments_evolution_exponential}
\end{figure}

For the numerical solution of the FPE (\eq \ref{eq:FPE_general}), we choose a finite element discretization method \cite{fenics_book} and evolve the system using either a stabilized Crank--Nicolson (CN) scheme \cite{knabner_angermann} in \fig \ref{fig:moments_evolution_power_law} or an Euler timestepping scheme \cite{fenics_book} in \fig \ref{fig:moments_evolution_exponential}.

The relative disagreement between MC simulations and finite-element solutions stays largely constant across different lags $k$ when we use the CN scheme instead of the Euler scheme as can be seen by comparing the lower panels of \figs \ref{fig:moments_evolution_power_law} and \ref{fig:moments_evolution_exponential}; the sizes of the disagreement are comparable in magnitude. This suggests that the remaining small discrepancy between MC simulations and PDE results can be largely explained with the errors associated with the MC simulation method. In particular, note that for the examples we show, the MC-GS weak error of size $\mathcal O(h)$ is comparable in magnitude to the numerator of $\Delta(Z)$ (\eq \ref{eq:relative_error}), i.e. the absolute disagreement. We will see what effects this has on the computation of correlations in Section \ref{sec:scc}.

\section{Transition to stationarity}

\begin{figure}
\centering
\includegraphics[width = 0.238\textwidth]{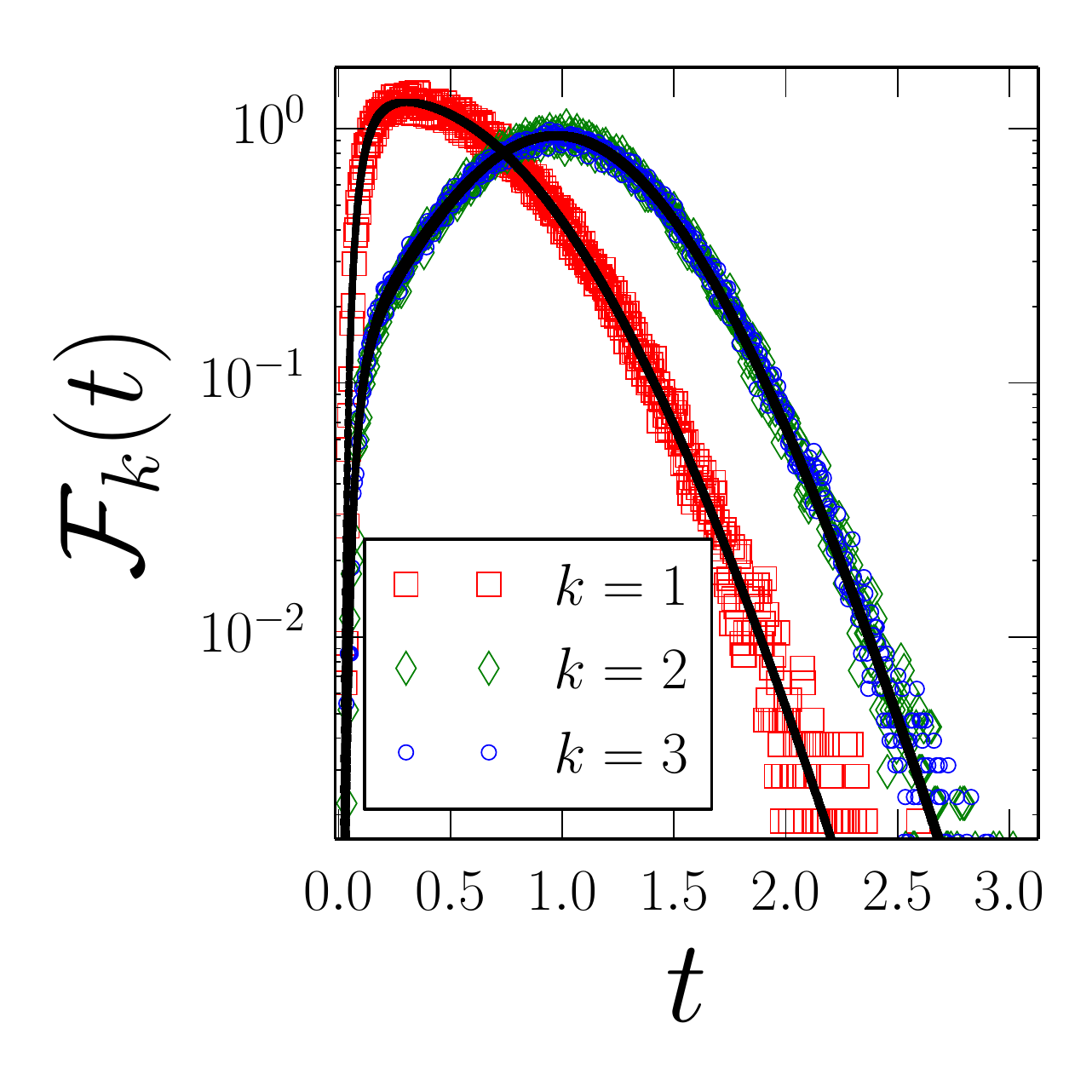}
\includegraphics[width = 0.238\textwidth]{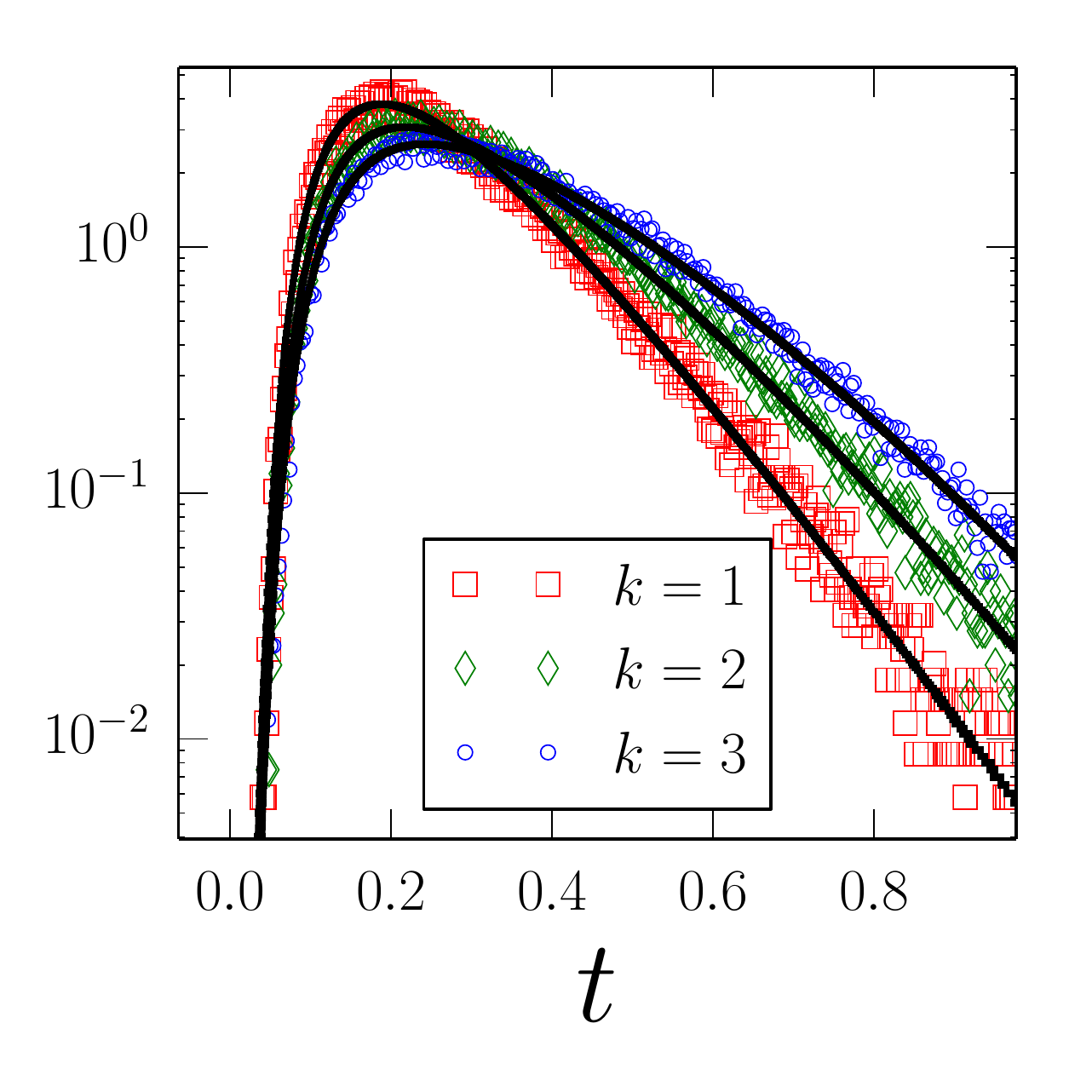}
\caption{PDF $\mathcal F_{k}$ of the $k$th ISI $T_{k}$ (\eq \ref{eq:kth_firing_time}). The symbols are MC simulations ($M = 10^{6}$ MC realizations) as indicated in the legends.  Solid black lines: PDF obtained by numerical solution of the FPE, \eq \ref{eq:FPE_general}. In the left panel, the distributions are practically indistinguishable after $k=2$. \textbf{Left:} Power-law adaptation (\eq \ref{eq:power_law}). \textbf{Right:} Single exponential adaptation (\eq \ref{eq:single_exponential}). Parameter values as in \fig \ref{fig:moments_evolution_power_law} for power-law adaptation and as in \fig \ref{fig:moments_evolution_exponential} for exponential adaptation. Both panels show results for plain MC simulations.}
\label{fig:histogram_fk}
\end{figure}

\begin{figure}
\centering
\includegraphics[width = 0.238\textwidth]{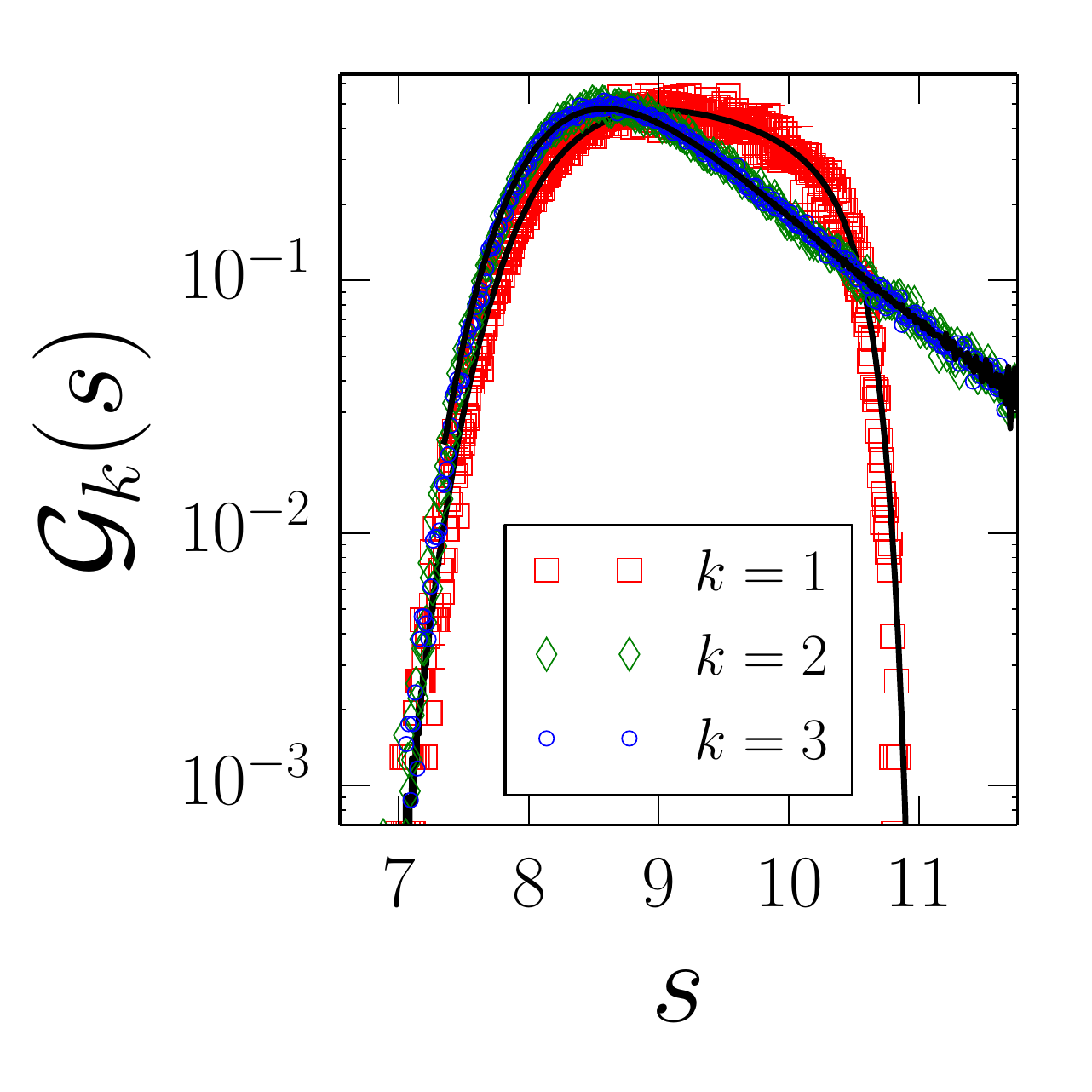}
\includegraphics[width = 0.238\textwidth]{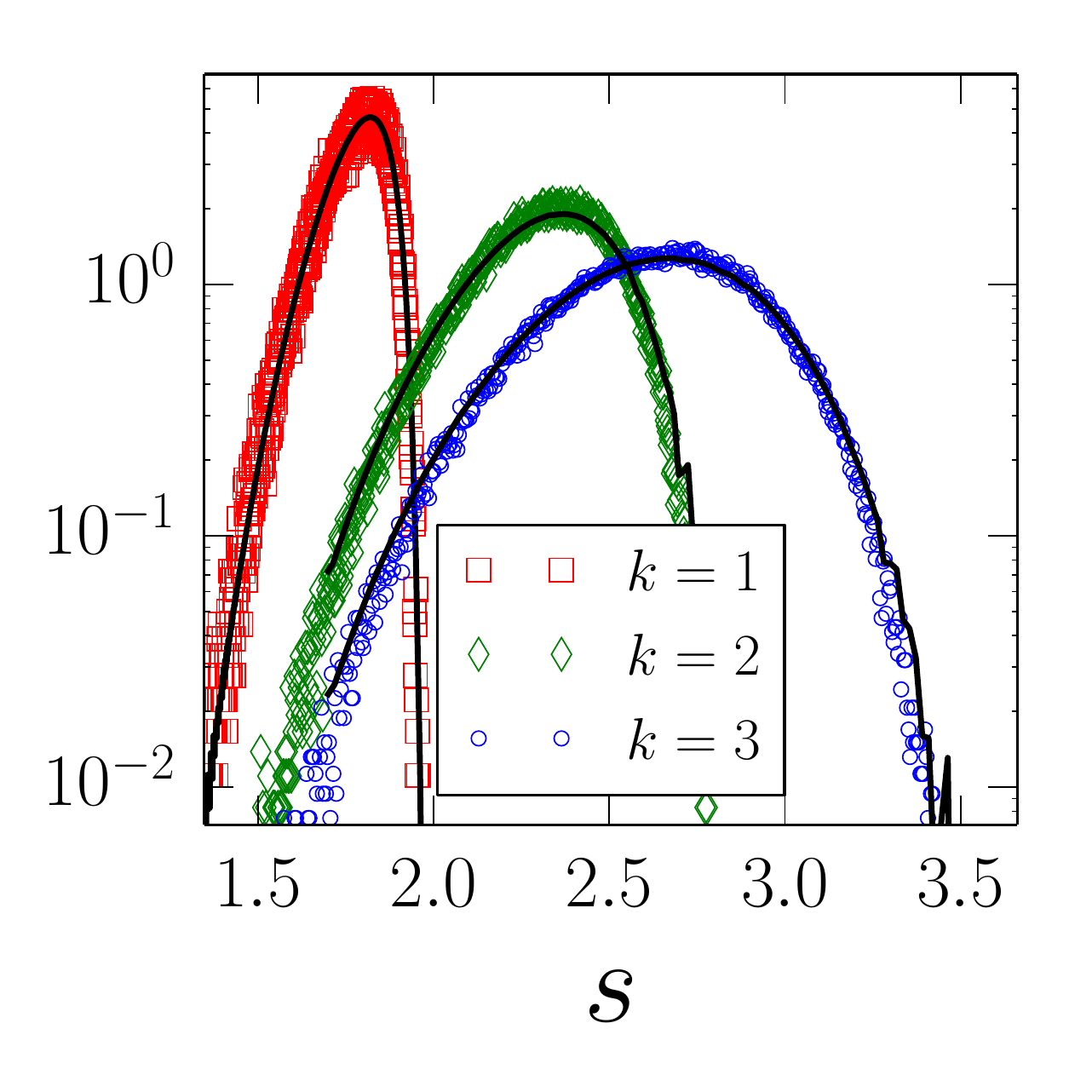}
\caption{PDF $\mathcal G_{k}$ of $s^{(k)}_{0}$ (\eq \ref{eq:kth_adaptation_current}). The symbols are MC simulations ($M = 10^{6}$ MC realizations) as indicated in the legends. The black solid line is $\mathcal G_{1}$ obtained from \eq \ref{eq:G_1} or  $\mathcal G_{2}$ ($\mathcal G_{3}$) obtained using \eq \ref{eq:density_z_integration}. In the left panel, the distributions are practically indistinguishable after $k=2$. \textbf{Left:} Power-law adaptation (\eq \ref{eq:power_law}). \textbf{Right:} Single exponential adaptation (\eq \ref{eq:single_exponential}). Parameter values as in \fig \ref{fig:moments_evolution_power_law} for power-law adaptation and as in \fig \ref{fig:moments_evolution_exponential} for exponential adaptation. Both panels show results for plain MC simulations.}
\label{fig:histogram_gk}
\end{figure}

We show the evolution of the rate and standard deviation of ISIs in \figs \ref{fig:moments_evolution_power_law} and \ref{fig:moments_evolution_exponential}. The rates decreases, whereas the standard deviation increases until both quantities reach a stationary value. Given that these quantities are derived from moments of the ISI distributions, the distributions also converge towards a stationary form. The convergence towards stationarity of ISI and peak adaptation current distributions ($\mathcal F_{k}$ and $\mathcal G_{k}$, respectively) is shown in \figs \ref{fig:histogram_fk} and \ref{fig:histogram_gk}. As expected for adapting models, the ISIs (whose distributions are shown in \fig \ref{fig:histogram_fk}) increase with higher $k$, which means that the rate $r_{k}$ decreases. This is well captured by the iFPT approach, with a maximal relative disagreement smaller than $3 \%$ in \fig \ref{fig:moments_evolution_power_law} and smaller than $2 \%$ in \fig \ref{fig:moments_evolution_exponential}. Also, the width of both the ISI distributions and the peak adaptation current distributions (\fig \ref{fig:histogram_gk}) increases, which is reflected by the increase of the variances of $\mathcal F_{k}$ and $\mathcal G_{k}$ shown in \fig \ref{fig:moments_evolution_power_law} and \ref{fig:moments_evolution_exponential}. The mean of the peak adaptation currents shifts to the right as stationarity is reached.
Moreover, stationarity is reached with varying speed, i.e. for different values of the lag $k$ (compare \fig \ref{fig:moments_evolution_power_law} with \fig \ref{fig:moments_evolution_exponential}). Generally, the speed of adaptation can be controlled by adjusting the bias current $I_{0}$ and the noise level $\sigma$ as well as the adaptation strength (size of the kick $\kappa$ and, in the case of single exponential adaptation, the timescale $\tau_{a}$). We have carried out additional MC simulations (not shown) to obtain insight into how these parameters influence the speed of the transition to stationarity. A higher bias current and a higher noise level will in general lead to a less rapid transition to stationarity.  This can be understood as follows: $X$ is driven to threshold more rapidly, causing the inhibition to build up quickly, reaching values that are higher than those typically found around the peak of the stationary distribution. This slows down the transition to stationarity, because $s$ needs to decay first. Moreover, a large kick size $\kappa$ and a rapidly decaying adaptation current will cause a quick transition. For the latter case, this is easily understood as we are then nearly dealing with a renewal system: after a short initial period, the effect of the adaptation current on the firing time statistics is negligible. For the former case, we note that the larger the kick size $\kappa$, the more pronounced the inhibitory effect of adaptation within one ISI, which means that $X$ takes longer to reach threshold before a large out-of-equilibrium average value of the current $s$ (a value that is larger than those typically found around the mode of the stationary distribution) can build up. The system reaches stationarity rapidly because it is quasi-deterministic as the dynamics of $s$ dominates the system, and the stochastic fluctuations of $X$ will only cause small perturbations. For both power-law and single exponential adaptation currents, it is possible to reach the stationary regime already after one or two firing events as in \fig \ref{fig:moments_evolution_power_law}, or to have a long transient regime as in \fig \ref{fig:moments_evolution_exponential}. The initial condition $s^{(0)}_{0}$ for the adaptation current can also be chosen to control the speed of the transition. If it is placed far away from the mean of the stationary distribution, the transition will take a longer time; also, it is possible to obtain a non-monotonic behaviour of the rate as a function of the interval number when $s^{(0)}_{0}$ is placed far above the aforementioned mean. The first mean ISI will then be the longest statistically, in contrast to the examples we show in \figs \ref{fig:moments_evolution_power_law} and \ref{fig:moments_evolution_exponential}.

\section{Statistics of the $k$th ISI}
\label{sec:gk_computation}

The iFPT approach can be iterated beyond the first two firing events to obtain the distribution for the third ISI, $\mathcal F_{3}(t)$. However, for the computation of the third ISI, no equation similar to \eq \ref{eq:G_1} can be used to obtain $\mathcal G_{2}$ because $s$ was started from a distribution to obtain $\mathcal F_{2}$. Indeed, for one fixed time $T_{2}$, there are many different starting values $s^{(1)}_{0}$ due to the stochastic dynamics of $X$. {Importantly, the ISI $T_{2}$ and the initial condition $s^{(1)}_{0}$ are not independent random variables (for a large value of $s^{(1)}_{0}$, a large ISI $T_{2}$ is more probable and vice versa) so that we can obtain the value that $s$ reaches after the second firing by the following observation (focusing on power-law adaptation): given that  $T_{2} = \lambda$ and $s^{(1)}_{0} = \nu$, we have $s^{(2)}_{0}  = \frac{1}{\frac{\lambda}{\alpha}+ \frac{1}{\nu}}  + \kappa$. We have included the jump of size $\kappa$ due to the definition of $s^{(2)}_{0}$ (see \fig \ref{fig:ensemble_sketch}). This emphasizes that once two values in the triplet $(T_{2}, s^{(1)}_{0}, s^{(2)}_{0})$ are fixed, the third one is determined. In the following, we again use $\lambda$ to denote a fixed FPT and $\nu$ to denote a fixed initial value of the adaptation current $s$. Analogously, we generally have for $s^{(k)}_{0}$: given $T_{k}=\lambda$ and  $s_{0}^{(k-1)}=\nu$, $s^{(k)}_{0}=f(\lambda,\nu)$ is determined. The function $f$ determining the subsequent value of the peak adaptation current given the previous ISI $\lambda$ and the previous peak value of the adaptation current $\nu$ reads for power-law adaptation (\eq \ref{eq:power_law}):

\begin{equation}
f(\lambda, \nu) = \kappa + \frac{1}{\frac{\lambda}{\alpha} + \frac{1}{\nu}} \, ,
\label{eq:f_definition}
\end{equation}

whereas for exponential adaptation (\eq \ref{eq:single_exponential}), we have 

\begin{equation}
 f(\lambda, \nu) = \kappa+ \nu \exp \left(-\frac{\lambda}{\tau_{a}} \right) \, .
\label{eq:f_definition_exponential}
\end{equation}

An alternative way is to fix the value of $s^{(k)}_{0}$ and then put a constraint on the time $T_{k}$ when $s_{0}^{(k-1)} $ is fixed: given $s_{0}^{(k-1)}=\nu$ and $s^{(k)}_{0} = \theta$, $T_{k} = h(\nu, \theta)$ is determined, where, for power-law adaptation, we have the ISI as a function of the previous and subsequent adaptation values: 

\begin{equation}
h(\nu, \theta) = \alpha \left(\frac{1}{\theta-\kappa} - \frac{1}{\nu} \right) \,.
\label{eq:h}
\end{equation}

The function $h$ is defined by solving the equation $f(\lambda, \nu) = \theta$ for $\lambda$. To actually compute the density $\mathcal G_{k}$, we need to relate the above observations to densities that we can compute with the iFPT approach. To that end, we now define the conditional density $\mathcal H$:

\begin{equation}
\mathcal H(\lambda, \nu) \d \lambda \equiv \mathbb P \left(T_{1} \in (\lambda, \lambda + \d \lambda)|s^{(0)}_{0} = \nu \right) \, .
\label{eq:H_definition}
\end{equation}

We have used $T_{1}$ and $s^{(0)}_{0}$ in the definition \eq \ref{eq:H_definition} to stress that, for the purpose of the computation of $\mathcal H$, we only need to solve the FPT problem for $T_{1}$ using different values of the initial condition $s^{(0)}_{0}$. It will become apparent below that we only need to compute $\mathcal H$ once, because it does not depend on the firing index $k$. For a fixed value of $\nu$, $\mathcal H(\lambda, \nu)$ is an FPT probability density. Our notation emphasizes that $\mathcal H$ is a function of two variables. $\nu$ sets the level of initial inhibition, i.e. the starting value of $s$. We show the function $\mathcal H$ for both power-law and exponential adaptation in \fig \ref{fig:H}. We see that with increasing starting value $\nu$, the mode of the FPT distribution shifts to larger times. For power-law adaptation, the shape of the FPT distributions does not change much, whereas for exponential adaptation, the distributions become broader with increasing $\nu$.



\begin{figure}
\centering
\includegraphics[width = 0.238\textwidth]{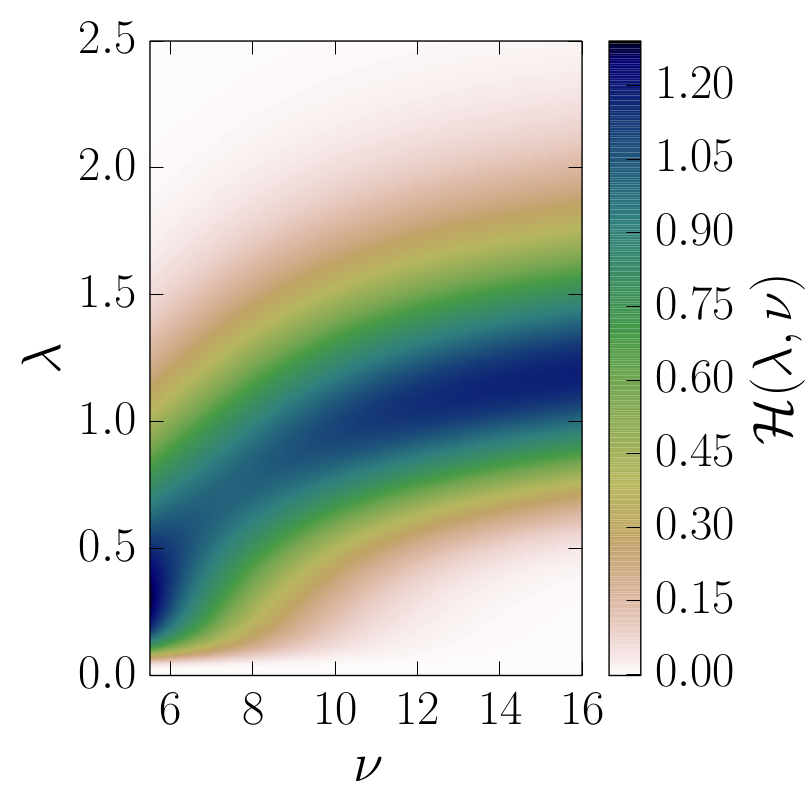}
\includegraphics[width = 0.238\textwidth]{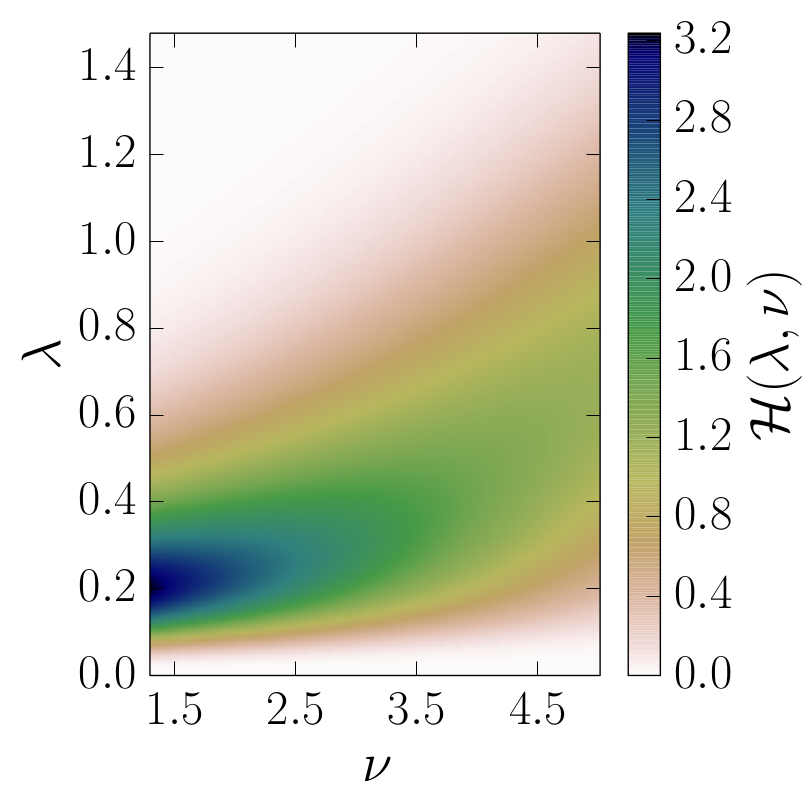}
\caption{Conditional FPT densities $\mathcal H(\lambda, \nu)$ given by \eq \ref{eq:H_definition} relating the initial value of the adaptation current to the distribution of the following ISI. Computed using numerical solutions to the FPE \eq \ref{eq:FPE_general}. \textbf{Left:} Power-law adaptation, computed using a CN timestepping scheme.  \textbf{Right:} Exponential adaptation, computed using an Euler timestepping scheme. Parameter values as in \fig \ref{fig:moments_evolution_power_law} for power-law adaptation and as in \fig \ref{fig:moments_evolution_exponential} for exponential adaptation.}
\label{fig:H}
\end{figure}

With this at hand, we now show how to practically compute the distributions $\mathcal G_{k}$ for $k>1$. We can obtain the CDF of $s_{0}^{(k)}$ by observing that:

\begin{equation}
\mathbb P\left(s^{(k)}_{0} \leq \theta \right) = \int_{\mathcal D^{(k-1)}(\theta)} \mathcal H(\lambda, \nu) \mathcal G_{k-1}(\nu) \d \lambda \d \nu \, ,
\label{eq:density_z_integration}
\end{equation}

with

\begin{equation}
\begin{split}
&\mathcal D^{(k-1)}(\theta) =\\
&\left( \lambda, \nu > 0 \arrowvert ~\nu \in \mathrm{supp}\left(\mathcal G_{k-1}\right),~h(\nu,\theta) \leq \lambda \leq T_{\max} \right) \, .
\end{split}
\label{eq:D_z_definition}
\end{equation}

The function $h$ defined in \eq \ref{eq:h} ensures that for a fixed value of $\nu$, we collect all times $\lambda$ so that $s_{0}^{(k)} \leq \theta$, which ensures that $f( \lambda, \nu ) \leq \theta$ for fixed values of $\theta$ and $\nu$. $T_{\max}$ is chosen so that $\mathcal H(T_{\max}, \nu) \approx 0~\forall \nu \in \mathrm{supp}\left(\mathcal G_{k-1} \right)$. This means that $T_{\max}$ should be chosen in the tail of the FPT distribution. Note that for the iFPT approach, one only has to compute $\mathcal H(\lambda,\nu)$ over the support of $\mathcal G_{k}$ for $k \geq 1$ once \footnote{The support of $\mathcal G_{k}$ is the open interval $(0, s_{0}^{(0)} + k \cdot \kappa)$.}, and then  multiply it with the adaptation current distribution of the previous iteration $\mathcal G_{k-1}$. This function then needs to be integrated according to \eq \ref{eq:density_z_integration}, and the PDF $\mathcal G_{k}$ can be obtained by numerical differentiation.
We show results for $\mathcal G_{2}$ and $\mathcal G_{3}$ using \eq \ref{eq:density_z_integration} in \fig \ref{fig:histogram_gk}. The agreement between MC simulations and \eq \ref{eq:density_z_integration} is excellent.

Iterating these ideas into the stationary regime, the ideas of the previous paragraph can be used to directly compute the stationary density of the peak adaptation current. We assume that stationarity is reached after $k^{\ast}-1$ firing events. Then  $s_{0}^{(k^{\ast})}$ and $s_{0}^{(k^{\ast} +1)}$ have the same distribution.  Consequently, the stationary density of the peak adaptation current after firing satisfies the two-dimensional integral equation (cf. \eq \ref{eq:density_z_integration})

\begin{equation}
\mathcal Q(\theta) = \int_{\mathcal D(\theta)} \mathcal H(\lambda, \nu) \mathcal Q^{\prime}(\nu) \d \lambda \d \nu \, ,
\label{eq:integral_equation_stationary_density_s}
\end{equation}

where $ \mathcal Q(\theta) \equiv \mathbb P \left(s^{(k^{\ast})}_{0} \leq \theta \right)$ denotes the CDF of the stationary peak value for the adaptation current $s$ (\eq \ref{eq:s_definition}). We have checked that this equation is indeed satisfied by the stationary distributions for the peak adaptation current obtained from MC simulations (data not shown). Therefore, \eq \ref{eq:integral_equation_stationary_density_s} can serve as a tool to check whether a given distribution for the peak adaptation current is stationary, or alternatively as a way to compute $\mathcal Q(\theta)$ directly if the function $\mathcal H$ is known.

\section{Correlations between Interspike Intervals} 
\label{sec:scc}

We now show how to compute serial correlations with the iFPT approach. We define the SCC \cite{urdapilleta_pre_onset_of_correlations_2011, mankin_rekker_2016} between the $n$th ISI $T_{n}$ and the $(n+k)$th ISI $T_{n+k}$ according to 

\begin{equation}
\mathrm{SCC}(n, k) = \frac{\mathbb E \left( T_{n} T_{n+k}\right) -\mathcal Q_{1}(n,k)}{\mathcal Q_{2}(n,k)} \, ,
\label{eq:SCC_definition_general}
\end{equation}

where

\begin{equation}
\mathcal Q_{1}(n,k) = \mathbb E (T_{n}) \mathbb E(T_{n+k}) \, ,
\label{eq:Q_1}
\end{equation}

and
\begin{equation}
\mathcal Q_{2}(n,k) = m_{2}(k)m_{2}(n+k) =  \sqrt{\mathrm{Var}(T_{n})\mathrm{Var}(T_{n+k})} \, .
\label{eq:Q_2}
\end{equation}

Here, $\mathrm{Var}(T_{n})$ denotes the variance of the $n$th ISI distribution, and $m_{2}$ is the standard deviation given by \eq \ref{eq:m2_definition}. Note that the definition \eq \ref{eq:SCC_definition_general} does not make use of the notion of stationarity, so that the SCC depends on both the position $n$ of the ISI in the spike train as well as on the lag $k$ between ISIs. Since we have already computed the distributions of the $k$th ISI, we can readily compute the variances and means in \eq \ref{eq:SCC_definition_general}, i.e. the terms given by \eqs \ref{eq:Q_1} and \ref{eq:Q_2}. It is slightly more complicated to compute the first term in the numerator, $\mathbb E \left( T_{n} T_{n+k}\right)$, because we need the joint density  $p_{2}(T_{n}, T_{n+k})$ of $T_{n}$ and $T_{n+k}$. In the present study, we focus on $k=1$. By definition, we have

\begin{align*}
\mathbb{E}(T_nT_{n+1})&=\int \mathrm{d} T_n\mathrm{d}T_{n+1}\,T_n\, T_{n+1}\, p_2(T_n,T_{n+1}),\\
&=\int \mathrm{d}T_n\,\mathrm{d}T_{n+1}\,T_n\,T_{n+1}p_1(T_{n+1}|T_n)\mathcal{F}_n(T_n),
\end{align*}

where as previously $\mathcal{F}_n(T_n)$ is the density of the $n$th ISI and $p_1(T_{n+1}|T_n)$ is the conditional density of $T_{n+1}$ given $T_n$. Because $T_{n+1}$ is statistically determined only by $s_{0}^{(n)}$, we can define this conditional density as

\begin{align*}
 p_1(T_{n+1}|T_n)&=\int\mathrm{d}y\,p(T_{n+1},y|T_n),
\end{align*}

where $p(T_{n+1},y|T_n)$ denotes the joint density of $T_{n+1}$ and
$s_{0}^{(n)}=y$ conditioned on the previous ISI $T_{n}$. We can rewrite this as follows:

\begin{align*}
  p(T_{n+1},y|T_n)&=\frac{p_{3}(T_{n+1},y,T_n)}{p(T_n)},\\
                  &=\frac{p(T_{n+1}|y,T_n)p(y,T_n)}{p(T_n)},\\
                  &=\frac{p(T_{n+1}|y,T_n)p(y|T_n)p(T_n)}{p(T_n)},\\
                  &=p(T_{n+1}|y,T_n)p(y|T_n).
\end{align*}
Now, as we have previously shown, the statistics of $T_{n+1}$ is completely determined when $s_0^{(n)}=y$ is fixed, hence $p(T_{n+1}|y,T_n)=p(T_{n+1}|y)\equiv \mathcal{H}(T_{n+1},y)$. Therefore, we have

\begin{align*}
\mathbb{E}(T_nT_{n+1})= \int \mathrm{d}T_n \, \mathrm{d}T_{n+1}\,\mathrm{d}y\,T_n\,T_{n+1}\mathcal{H}(T_{n+1},y)p(y|T_n)\mathcal{F}_n(T_n) \, .
\end{align*}

This can be further simplified by noting that $p(y|T_{n}) = \frac{p(y, T_{n})}{\mathcal F_{n}(T_{n})}$ and therefore

\begin{align}
\label{eq:n1_correct}
\mathbb{E}(T_nT_{n+1})=\int \mathrm{d} T_n\,\mathrm{d}T_{n+1}\,\mathrm{d}y\,T_n\,T_{n+1}\,\mathcal{H}(T_{n+1},y)p(y,T_n) \, .
\end{align}

For $n=1$, \eq \ref{eq:n1_correct} can be simplified because $s_{0}^{(1)}$ is a deterministic function of $T_{1}$ (see \eq \ref{eq:G_1}), so that $p(s_{0}^{(1)} = y, T_{1} = x) = \delta\left(y-f(x, s_{0}^{(0)})\right) \mathcal F_{1}(x)$, where $f$ is defined by \eqs \ref{eq:f_definition} and \ref{eq:f_definition_exponential}. Hence, we have for $n=1$

\begin{equation*}
\mathbb{E}(T_1T_{2})=\int \mathrm{d} T_1 \, \mathrm{d}T_{2}\, T_1\, T_{2}\,\mathcal{H}(T_{2},f(T_{1}, s_{0}^{(0)}))\mathcal F_{1}(T_{1}).
\end{equation*}

We have shown in Section \ref{sec:gk_computation} how to obtain the conditional FPT density $\mathcal H$. To evaluate \eq \ref{eq:n1_correct} for general $n$, we still need to compute the joint density $p\left(s_{0}^{(n)} = y,T_{n}\right)$. This can be achieved by means of an MC simulation, where we fix a value of $n$ and then record the frequency with which pairs of $s_{0}^{(n)}$ and $T_{n}$ are generated by the system. We show an example of these densities in \fig \ref{fig:jd}. The most notable feature is an inverse proportionality between $s_{0}^{(n)}$ and $T_{n}$. The longer e.g. $T_{2}$, the less likely it is for the value of $s$ after the second firing, $s_{0}^{(2)}$, to attain a high value.



\begin{figure}
\centering
\includegraphics[width = 0.238\textwidth]{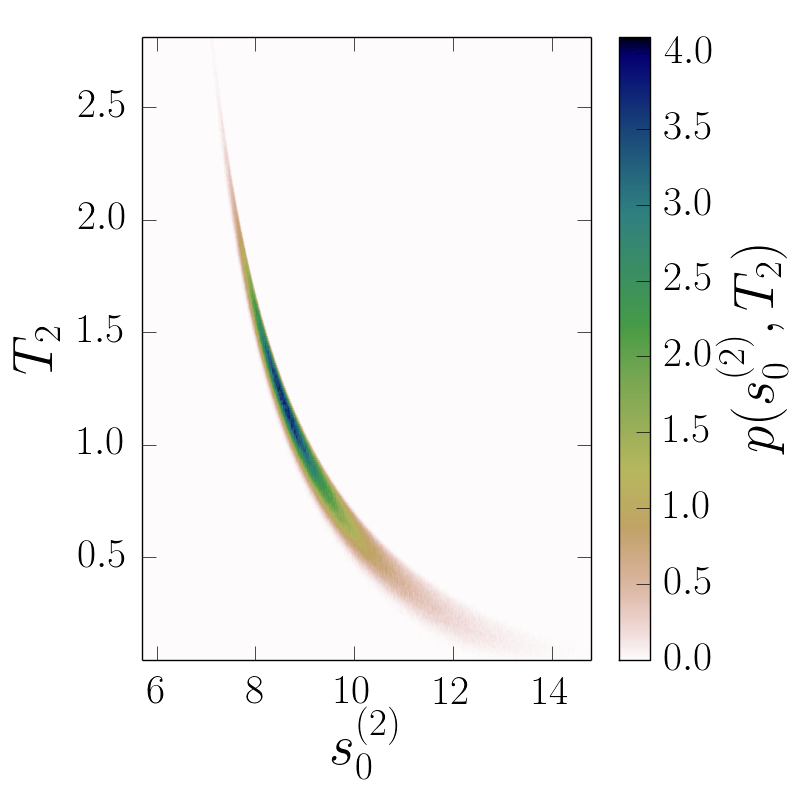}
\includegraphics[width = 0.238\textwidth]{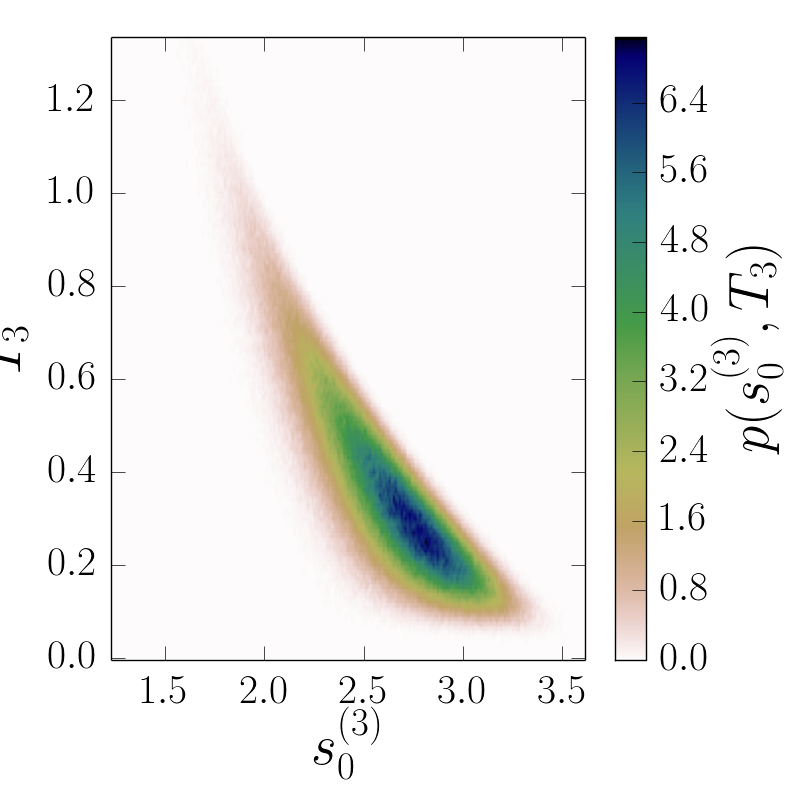}
\caption{MC simulations with GS boundary correction for the joint density of $s_{0}^{(n)}$ (i.e. $n$th peak value of adaptation current following the $n$th FPT) and $T_{n}$ for different values of $n$.  \textbf{Left:} Power-law adaptation, $n=2$.  \textbf{Right:} Exponential adaptation, $n= 3$. $M = 10^{6}$ MC realizations. Parameter values as in \fig \ref{fig:moments_evolution_power_law} for power-law adaptation and as in \fig \ref{fig:moments_evolution_exponential} for exponential adaptation.}
\label{fig:jd}
\end{figure}

\eq \ref{eq:n1_correct} is formally correct, but not very practical for actual computations. This is because to apply the iFPT approach, it is desirable to obtain all quantities needed for the SCC using solutions of the FPE only, and no MC simulations. These, however, are required to obtain an approximation for the joint density $p(s_{0}^{(n)}, T_{n})$ in \eq \ref{eq:n1_correct}. We therefore propose an approximation to compute $\mathbb E \left( T_{n} T_{n+1}\right)$ using the available densities $\mathcal F$, $\mathcal G$ and $\mathcal H$ only. To that end, we note that

\begin{equation}
\begin{split}
& p_{3}(T_{n+1}, T_{n}, y) = p(T_{n+1},T_{n} |y) p(y) \, .
\end{split}
\end{equation}

If we now assume that $T_{n}$ and $s_{0}^{(n)}$ are independent, we can approximate this as follows:

\begin{equation}
\begin{split}
&p_{3}(T_{n+1}, T_{n}, y)  \approx p(T_{n+1} |y) p(y) p(T_{n}) = \\
&\mathcal H(T_{n+1},y) \mathcal G_{n}(y) \mathcal F_{n}(T_{n}) \, .
\end{split}
\end{equation}

This results in an alternative, approximative expression for the expectation $\mathbb E\left(T_{n}T_{n+1}\right)$:

\begin{equation}
\begin{split}
&\mathbb E \left( T_{n} T_{n+1}\right) = \\
&\int\d T_{n} \d T_{n+1} \d y\, T_{n}\, T_{n+1}\, \mathcal H(T_{n+1}, y) \mathcal G_{n}(y) \mathcal F_{n}(T_{n})  \, .
\end{split}
\label{eq:e_n1_final}
\end{equation}

\eq \ref{eq:e_n1_final} is therefore equivalent to \eq \ref{eq:n1_correct} if $p(y, T_{n}) = \mathcal G_{n}(y) \mathcal F_{n}(T_{n})$. Although \fig \ref{fig:jd} demonstates that $p$ does not factorise (the joint density is negatively sloped), we show below that \eq \ref{eq:e_n1_final} approximates \eq \ref{eq:n1_correct} very well. Given that \eq \ref{eq:e_n1_final} does not require additional MC simulations, the small error introduced by \eq \ref{eq:e_n1_final} is well offset by the large reduction in computational cost.

There exists a third alternative expression for the expectation of the product of ISIs suggested for a different, but related, model, in \cite{chacron_pakdaman_longtin_2003}. It reads in our notation \footnote{Note that \eq 3.17 in \cite{chacron_pakdaman_longtin_2003} is in the stationary state: $ \mathbb E \left( T_{n} T_{n+1}\right) = \int\d T_{n} \d T_{n+1} \d y~ T_{n} T_{n+1} \mathcal H(T_{n+1}, f(T_{n},y))\mathcal H\left(T_{n},y \right) \mathcal G^{\ast}(y)$.}

\begin{equation}
\begin{split}
&\mathbb E \left( T_{n} T_{n+1}\right) = \\
&\int\d T_{n} \,\d T_{n+1}\,\d y \, T_{n}\,T_{n+1}\,\mathcal H(T_{n+1}, f(T_{n}, y)) \mathcal H(T_{n}, y) \mathcal G_{n-1}(y) \, ,
\end{split}
\label{eq:n1_traditional}
\end{equation}

where $f$ is given by \eq \ref{eq:f_definition} or \eq \ref{eq:f_definition_exponential}. The term $\mathcal H(T_{n}, y) \mathcal G_{n-1}(y)$ that appears in \eq \ref{eq:n1_traditional} is the same as the one on the right-hand side in \eq \ref{eq:density_z_integration}, which we used to obtain the CDF of $s_{0}^{(n)}$.

For $n=1$, \eq \ref{eq:n1_traditional} reads

\begin{equation*}
\begin{split}
&\mathbb E \left( T_{1} T_{2}\right)=\\
&\int\d T_{1}\,\d T_{2}\,T_{1}\,T_{2}\,\mathcal H\left(T_{2}, f\left( T_{1}, s_{0}^{(0)}\right)\right) \underbrace{\mathcal H\left(T_{1}, s_{0}^{(0)}\right)}_{=\mathcal F_{1}(T_{1})} \, ,
\end{split}
\label{eq:n1_traditional_1}
\end{equation*}

because $s$ is started from a point $s_{0}^{(0)}$, so that formally $\mathcal G_{0}(y) = \delta \left(y-s_{0}^{(0)}\right)$, which collapses the integration over $y$ in \eq \ref{eq:n1_traditional}. Thus, for $n=1$, \eq \ref{eq:n1_correct} and \ref{eq:n1_traditional} coincide. This is also true for higher values of $n$. A proof for this equivalence is presented in Appendix \ref{sec:appendix}. \eq \ref{eq:n1_traditional} only makes use of the quantities $\mathcal H$ and $\mathcal G$, which can be computed using the iFPT approach as explained in the previous section.

We show comparisons between MC simulations and the three expressions for $\mathbb E\left(T_{n}T_{n+1} \right)$, \eq \ref{eq:n1_correct}, \eq \ref{eq:e_n1_final} and \eq \ref{eq:n1_traditional}, in \fig \ref{fig:e_nk}. The results presented in \fig \ref{fig:e_nk} are in agreement with the observation that the two expressions \eq \ref{eq:n1_correct} and \eq \ref{eq:n1_traditional} are equivalent. We find that the agreement of \eq \ref{eq:n1_traditional} with MC simulations is comparable to \eq \ref{eq:e_n1_final}, particularly for exponential adaptation. Interestingly, the formally correct \eq \ref{eq:n1_correct} and the approximate \eq \ref{eq:e_n1_final} give comparable results; \eq \ref{eq:n1_correct} slightly deviates from MC simulations and \eq \ref{eq:e_n1_final} when $n$ gets larger. The maximal relative disagreement between MC and iFPT results is less than $2 \%$  (\fig \ref{fig:e_nk}, bottom panels). We will see below that the SCC is best approximated by using the exact result \eq \ref{eq:n1_correct} (or equivalently \eq \ref{eq:n1_traditional}), as we expect. We attribute the discrepancy between MC simulations and the exact result \eq \ref{eq:n1_correct} to the error caused by the numerical integration over the MC approximation of the joint density $p(y, T_{n})$. We checked that applying a kernel density estimation \cite{scipy_kde} to the MC results for $p(y, T_{n})$ did not alter these results. 

Similar results for $\mathcal Q_{1}$ and $\mathcal Q_{2}$ (\eqs \ref{eq:Q_1} and \ref{eq:Q_2}) are shown in \figs \ref{fig:Q1_nk} and \ref{fig:Q2_nk}. The agreement is good, with the maximal relative disagreement always less than $5 \%$. The relative disagreement for the statistics of the product of two adjacent ISIs, $\mathcal Q_{1}(n,1)$, is in general larger than the error for the moments, as can be seen by comparing \figs \ref{fig:moments_evolution_power_law} and \ref{fig:moments_evolution_exponential} with \fig \ref{fig:Q1_nk}. Indeed, for the case of power-law adaptation, we observe an increase of roughly one order of magnitude in the relative error even when the more accurate CN scheme is used (see e.g. left panels of \fig \ref{fig:Q1_nk}). An exception is the computation of the joint expectation, shown in \fig \ref{fig:e_nk}, where, depending on which methods are compared, the relative disagreement is comparable in size to the one for the computation of the moments shown in \figs \ref{fig:moments_evolution_power_law} and \ref{fig:moments_evolution_exponential}. This increase of the relative disagreement makes the computation of the SCC using the iFPT approach hard, because the two expressions in the numerator of \eq \ref{eq:SCC_definition_general} are quite close to one another for the parameter values we have chosen here, meaning that the numerator is small and indeed of the same magnitude or even smaller as the relative disagreement, e.g. $-2.6\cdot 10^{-2}$ in the left panel and $-8.6 \cdot 10^{-4}$ in the right panel of \fig \ref{fig:e_nk} for $n=1$ for the MC-GS simulation method. The increase in the relative discrepancy is caused by error propagation, because for the second-order statistics, one has to multiply two quantities that both come with an individual error. Therefore, whereas the iFPT approach can in principle also be used to compute serial correlations present in the spike train, obtaining reliable results can in general be a computational challenge. When the negative serial correlations are stronger, so that the difference in the numerator of \eq \ref{eq:SCC_definition_general} is larger, the iFPT approach should give more accurate results. We stress that the dominant source of error is not the computation of the joint expectation $\mathbb E\left(T_{n}T_{n+1}\right)$ of ISIs, but the product of the expectation of ISIs and the variances, which can be seen by comparing the lower panels of \fig \ref{fig:e_nk} with those of \figs \ref{fig:Q1_nk} and \ref{fig:Q2_nk}. 

Finally, we show the SCC at lag $1$ obtained by MC simulations and PDE numerics in \fig \ref{fig:SCC_MC}. The agreement is worse than for all previously considered quantities, but still reasonable. To verify the MC simulations, we checked that our MC simulation setup was able to reproduce known analytical results for the SCC obtained in \cite{urdapilleta_pre_onset_of_correlations_2011} for certain limiting cases.  The anti-correlations between adjacent ISIs ($\mathrm{SCC}(n,1) <0$) strengthen until they reach a stationary value. 

Thus, we see that MC and PDE results for the SCC in general do not agree as well as one would expect from the good agreement of the expecations $\mathbb E(T_{n}T_{n+1})$ in \fig \ref{fig:e_nk}.  The deviation is likely more  pronounced for parameters that lead to small negative SCCs, which we have for both models considered in this section. In the next Section, we will compare this with results for the perfect integrate-and-fire model, where parameter values are chosen so that the SCCs are more negative and hence the agreement is better. This is because the two terms in the numerator of \eq \ref{eq:SCC_definition_general} are close to each other for small SCCs, and hence a small error in them impacts the accuracy of the SCC computation quite dramatically.



\begin{figure}
\centering

\includegraphics[width = 0.5\textwidth]{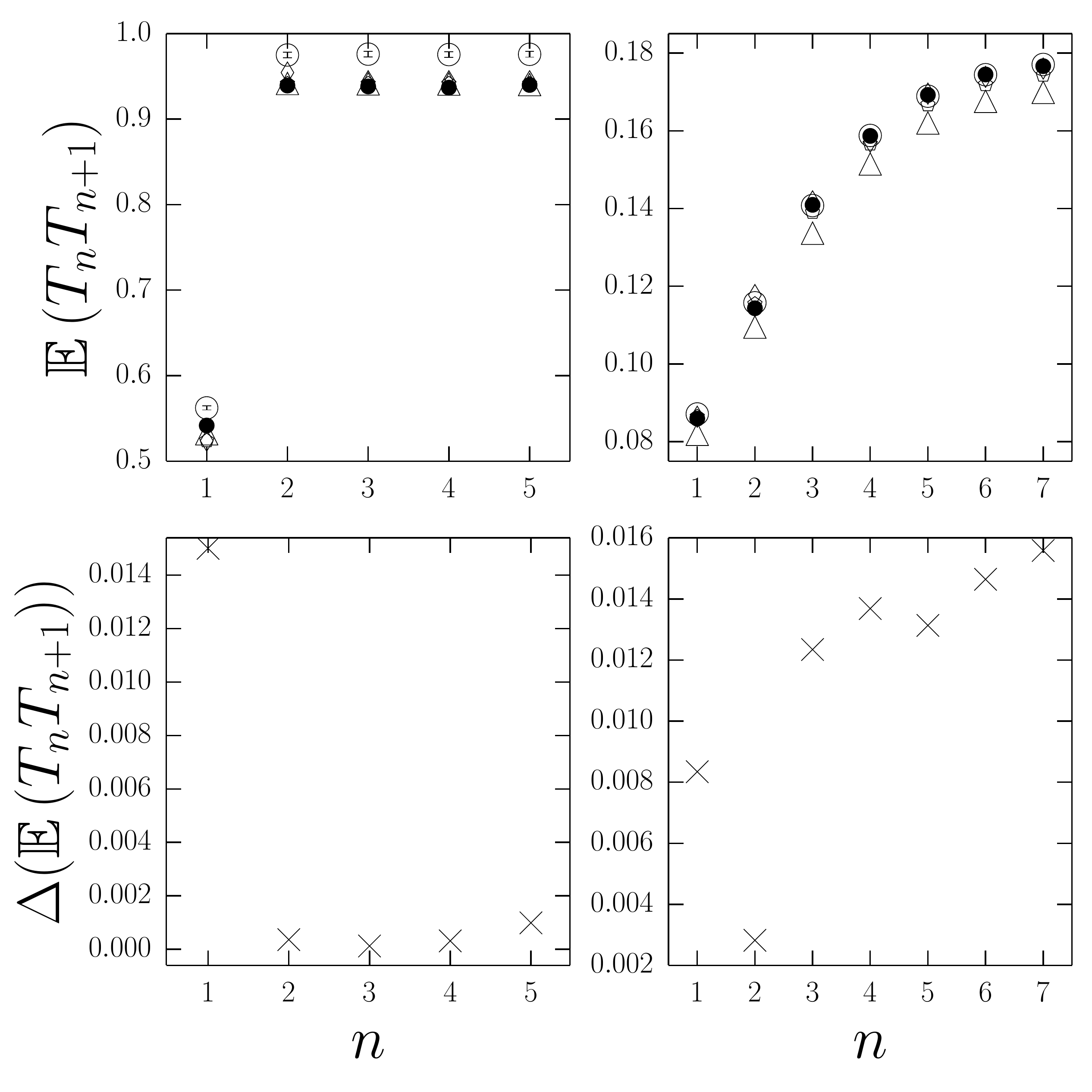}
\caption{\textbf{Top:} Expectation $\mathbb E \left(T_{n} T_{n+1} \right)$ of adjacent ISIs for different values of $n$. Left: power-law adaptation. Right: exponential adaptation. $M = 10^{6}$ MC realizations. Empty circles: Plain MC simulations of \eqs \ref{eq:X_definition} and \ref{eq:s_definition}. Triangles: MC simulations with GS boundary correction. Pentagons: \eq \ref{eq:n1_correct}. For power-law adaptation, the pentagons are on top of the empty triangles and filled circles and hence not visible. Filled circles: \eq \ref{eq:e_n1_final}. Diamonds: \eq \ref{eq:n1_traditional}. The diamonds are nearly  on top of the filled circles and hence not visible. The PDE results were obtained using a CN scheme (power-law adaptation) or an Euler timestepping scheme (exponential adaptation). The vertical error bars show the MC error for a $>99.99 \%$ confidence interval. \textbf{Bottom:} Relative disagreements defined by \eq \ref{eq:relative_error}, where for power-law adapation, the GS boundary corrected MC algorithm was used, and for exponential adaptation, the plain MC algorithm was used. For the iFPT quantity, \eq \ref{eq:n1_correct} was used. Parameter values as in \fig \ref{fig:moments_evolution_power_law} for power-law adaptation and as in \fig \ref{fig:moments_evolution_exponential} for exponential adaptation.}
\label{fig:e_nk}
\end{figure}
\begin{figure}
\centering
\includegraphics[width = 0.5\textwidth]{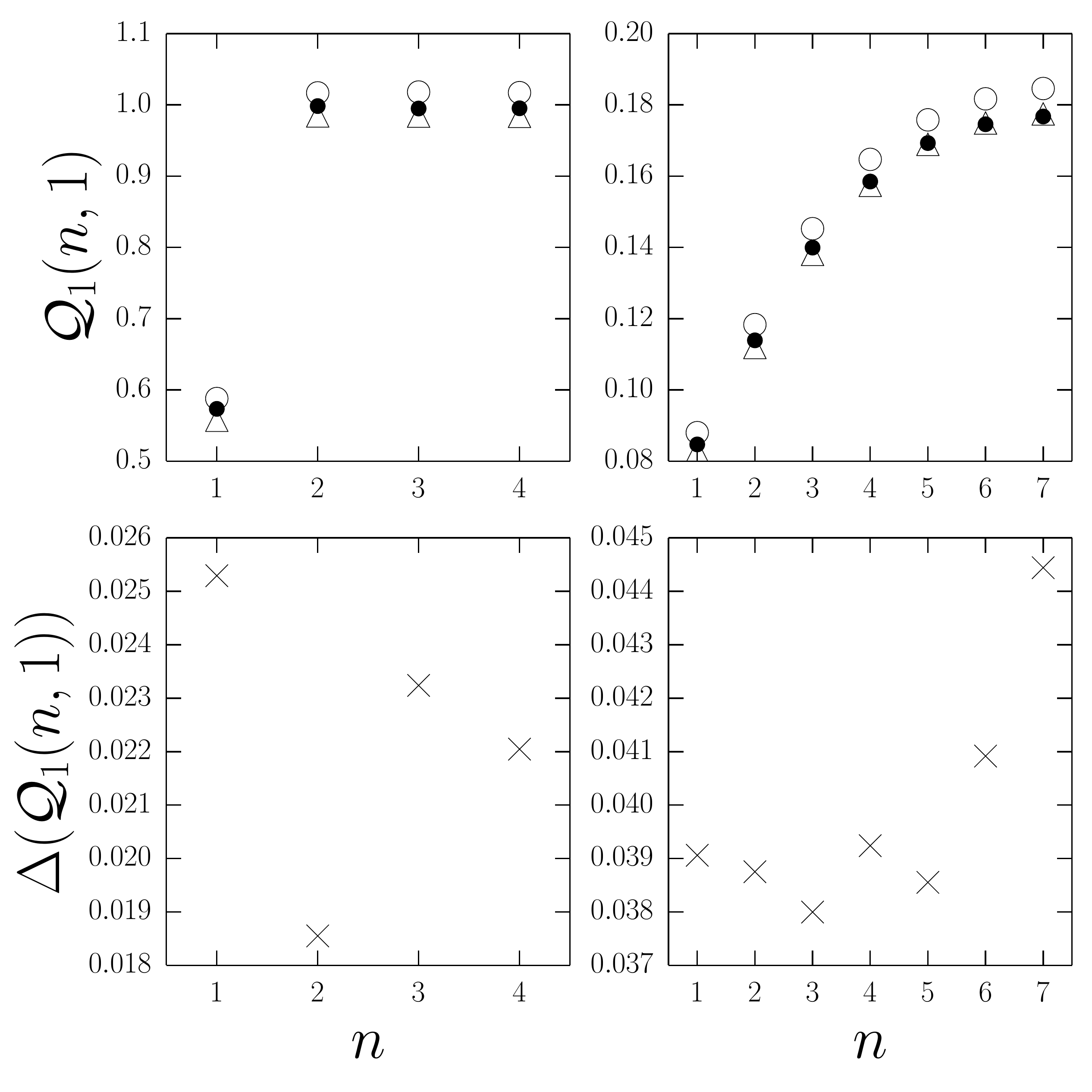}
\caption{\textbf{Top:} $\mathcal Q_{1}(n,1)$ (defined in \eq \ref{eq:Q_1}) of adjacent ISIs for different values of $n$. Left: power-law adaptation. Right: exponential adaptation. Empty circles: MC simulations of \eqs \ref{eq:X_definition} and \ref{eq:s_definition}. Triangles: MC simulations with GS boundary correction. Filled circles: \eq \ref{eq:Q_1}, CN timestepping scheme for power-law adaptation and Euler timestepping scheme for exponential adaptation. \textbf{Bottom:} Relative disagreement defined by \eq \ref{eq:relative_error}, a plain MC algorithm was used to obtain the relative disagreement. Parameter values as in \fig \ref{fig:moments_evolution_power_law} for power-law adaptation and as in \fig \ref{fig:moments_evolution_exponential} for exponential adaptation.}
\label{fig:Q1_nk}
\end{figure}

\begin{figure}
\centering
\includegraphics[width = 0.5\textwidth]{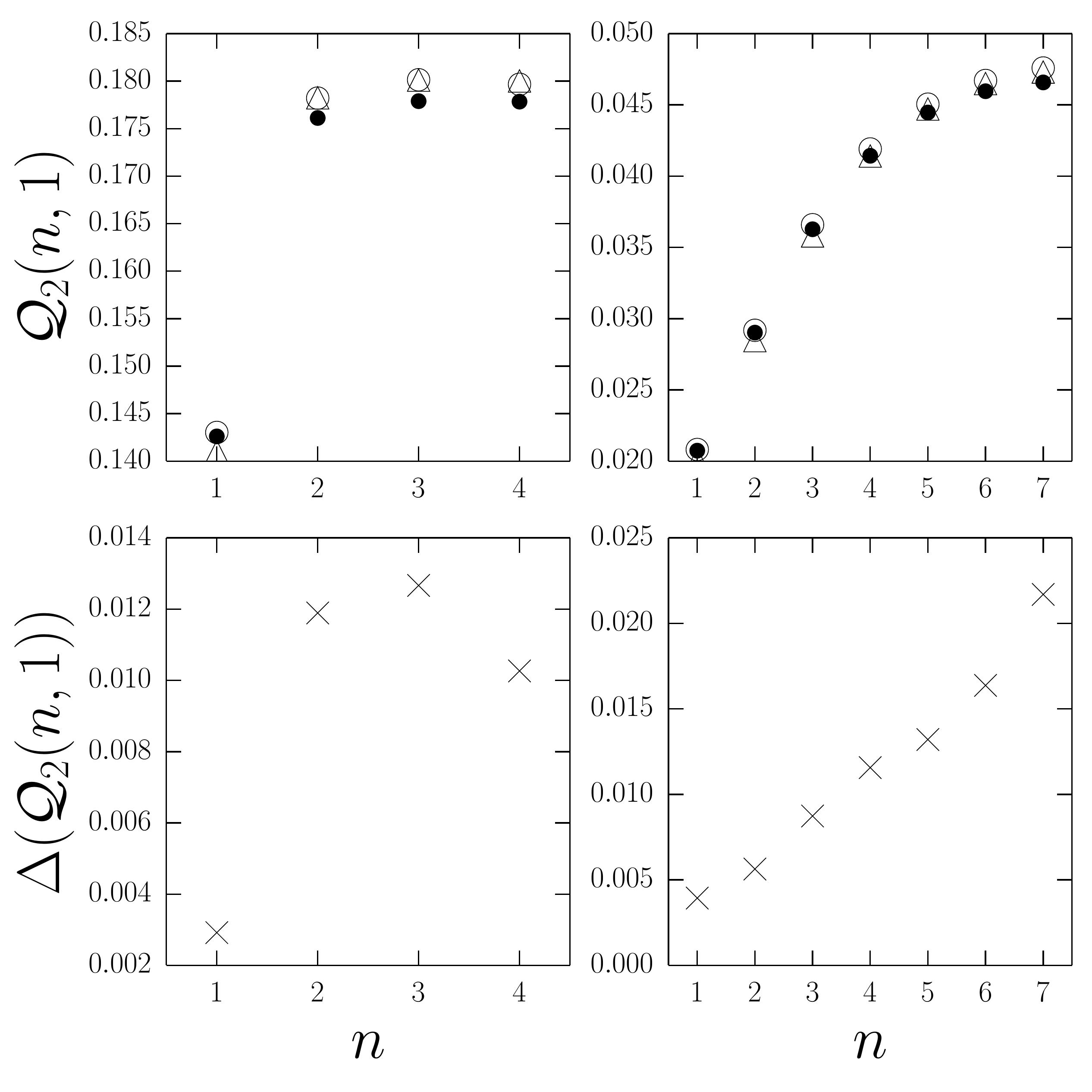}
\caption{\textbf{Top:} $\mathcal Q_{2}(n,1)$ (defined in \eq \ref{eq:Q_2}) for adjacent ISIs for different values of $n$. Left: power-law adaptation. Right: exponential adaptation. Empty circles: MC simulations of \eqs \ref{eq:X_definition} and \ref{eq:s_definition}. Triangles: MC simulations with GS boundary correction. Filled circles: \eq \ref{eq:Q_2}, CN timestepping scheme for power-law adaptation and Euler timestepping scheme for exponential adaptation. \textbf{Bottom:} Relative disagreement defined by \eq \ref{eq:relative_error}, a plain MC algorithm was used to obtain the relative disagreement. Parameter values as in \fig \ref{fig:moments_evolution_power_law} for power-law adaptation and as in \fig \ref{fig:moments_evolution_exponential} for exponential adaptation. Parameter values as in \fig \ref{fig:moments_evolution_power_law} for power-law adaptation and as in \fig \ref{fig:moments_evolution_exponential} for exponential adaptation.}
\label{fig:Q2_nk}
\end{figure}

\begin{figure}
\centering
\includegraphics[width = 0.5\textwidth]{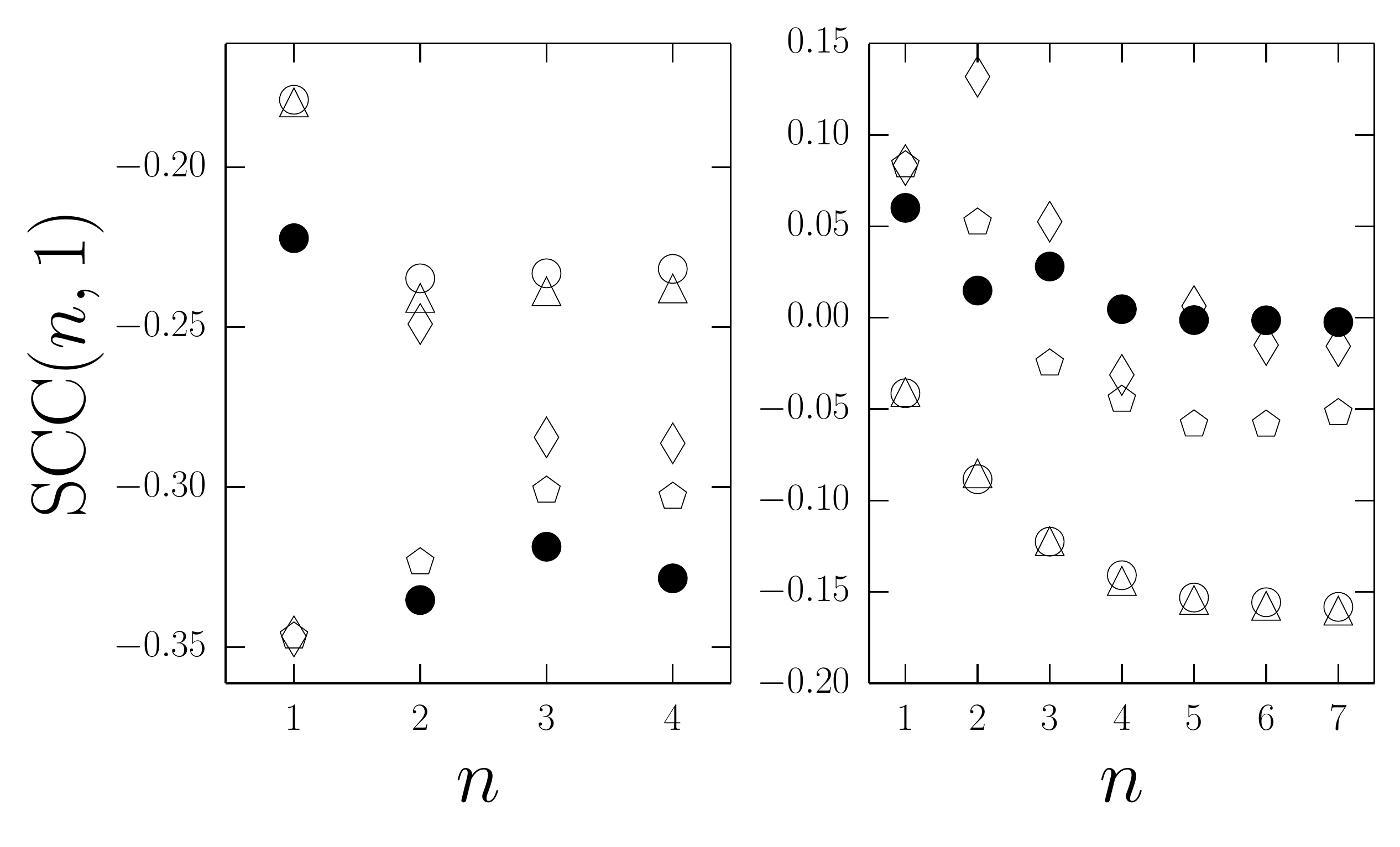}
\caption{ Serial correlation coefficient at lag $n=1$ (defined in \eq \ref{eq:SCC_definition_general}) for adjacent ISIs for different values of $n$ for the LIF model (\eq \ref{eq:X_definition}). \textbf{Left:} power-law adaptation. \textbf{Right:} exponential adaptation. Empty circles: Plain MC simulations of \eqs \ref{eq:X_definition} and \ref{eq:s_definition}. Triangles: MC simulations with GS boundary correction. Filled circles: PDE results, SCC computed using \eq \ref{eq:e_n1_final}. Pentagons: PDE results, SCC computed using \eq \ref{eq:n1_correct}. CN timestepping scheme for power-law adaptation and Euler timestepping scheme for exponential adaptation.  Diamonds: PDE results, SCC computed using \eq \ref{eq:n1_traditional}. Parameter values as in \fig \ref{fig:moments_evolution_power_law} for power-law adaptation and as in \fig \ref{fig:moments_evolution_exponential} for exponential adaptation. Even if the joint expectations shown in \fig \ref{fig:e_nk} agree well, this does not imply that the SCC will be well approximated; the small correlation values (i.e. the difference between $\mathbb E(T_{n}T_{n+1})$ and $\mathcal Q_{1}(n,1)$) for the two examples lead to large discrepancies in the SCCs, which are especially significant for the case of power-law adaptation.}
\label{fig:SCC_MC}
\end{figure}

We show in the next section that our methods reproduce known stationary analytical results for the SCC when we consider the perfect integrate-and-fire model with single exponential adaptation in a parameter regime where we have large negative correlations, thus demonstrating that our methods are sound, but SCC calculations are very sensitive to numerical inaccuracies.

\subsection{The perfect integrate-and-fire model}

The adapting perfect integrate-and-fire (PIF) model driven by white Gaussian noise and a single exponential adaptation current is one of the simplest models for spike-triggered adaptation. For small noise intensity, analytical expressions for the stationary SCC exist. We here study this stationary limit case and compare analytical formulas to results obtained with the iFPT approach. 

The model reads (we follow the notation of \cite{schwalger_13} and \cite{schwalger_lindner_2013})

\begin{align}
\d X &= (I_{0} -s) \d t + \sqrt{2D} \d W(t) \, ,\\
\frac{\d s}{\d t} &= -\frac{s}{\tau_{a}} \, .
\label{eq:PIF_definition}
\end{align}

The adaptation mechanism works in analogy to the previous model (\eq \ref{eq:s_definition}): whenever $X$ reaches the threshold $X = 1$, $s$ receives a kick of size $\Delta \equiv \frac{\widetilde{\Delta}}{\tau_{a}}$ and $X$ is instantaneously reset to $0$. 

The stationary SCC at lag $1$ for this model under the assumption of small noise (i.e. $D \ll 1$) reads \cite{schwalger_13}

\begin{equation}
 \mathrm{SCC}(k=1) = -\frac{\alpha(1-\theta)(1-\alpha^{2}\theta)}{1 + \alpha^{2} - 2\alpha^{2}\theta} \, ,
 \label{eq:scc_schwalger}
\end{equation}

where

\begin{align*}
\alpha  &= \frac{s^{\ast} - \Delta}{s^{\ast}}, \qquad \theta = \frac{I_{0} - s^{\ast}}{I_{0} - s^{\ast} + \Delta} \, ,\\
T^{\ast} &= \frac{1 + \widetilde{\Delta}}{I_{0}}, \qquad s^{\ast} = \frac{\Delta}{1 - \exp\left(-\frac{T^{\ast}}{\tau_{a}} \right)} \, .
\end{align*}

Thus, we can compute the SCC in closed analytical form as a function of the system parameters. This formula serves as an important benchmark for our numerical results. In particular, we expect that after the described transition to stationarity, the SCC given by \eq \ref{eq:SCC_definition_general} will approach the stationary SCC given by \eq \ref{eq:scc_schwalger}. This is confirmed in \fig \ref{fig:SCC_MC_PIF}. In particular, the agreement between MC simulations and the exact formula \eq \ref{eq:n1_correct} is very good (the relative disagreement between PDE numerics and the analytical result is less than $6 \%$ for the stationary value); the agreement of MC simulations with the approximation \eq \ref{eq:e_n1_final} is a bit worse, but still reasonable. Thus, we conclude that our methodology can be used more generally to compute the evolution of the moments and SCCs. However, as seen in the previous section, to obtain a good agreement between MC simulations and PDE numerics, the computational effort might be rather large. In particular, we note that the PIF example shown in \fig \ref{fig:SCC_MC_PIF} gives rise to stronger negative SCCs, which means that the error propagation has less of an effect, but is still present, even when moments of firing times between MC and PDE numerics disagree by less than $1 \%$ (data not shown). We finally note that it is also possible to analytically compute the stationary SCC at higher lags and for different models (e.g. the leaky integrate-and-fire model in the presence of weak noise or for small adaptation currents) using the approach described in \cite{schwalger_lindner_2013}, or, using a different approach, in \cite{urdapilleta_pre_onset_of_correlations_2011}.

\begin{figure}
\centering
\includegraphics[width = 0.43\textwidth]{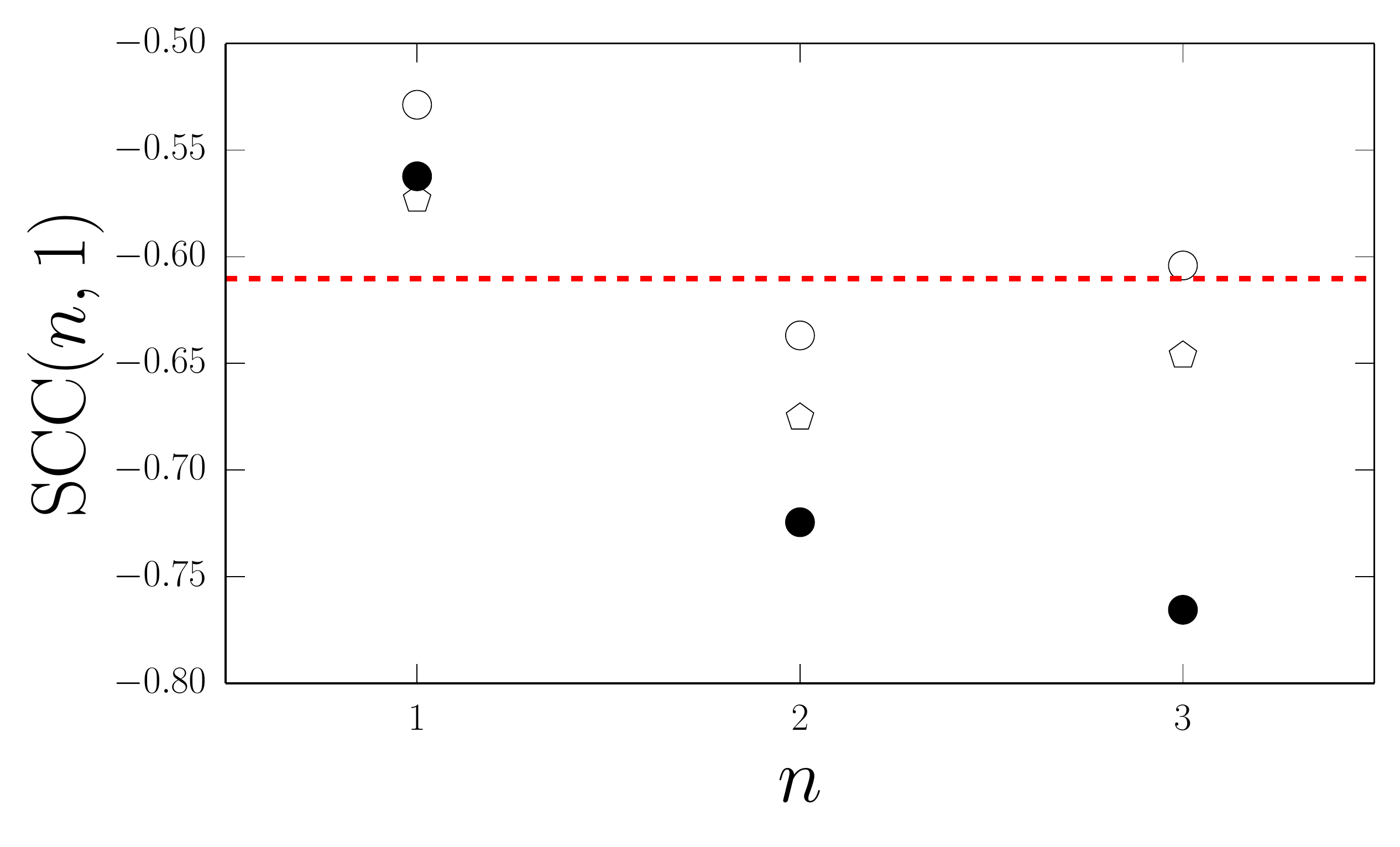}
\caption{ Serial correlation coefficient at lag $n=1$ (defined in \eq \ref{eq:SCC_definition_general}) for different values of $n$ for the PIF model (\eq \ref{eq:PIF_definition}). The dashed horizontal line is the stationary SCC given by \eq \ref{eq:scc_schwalger}. Empty circles: Plain MC simulations of \eq \ref{eq:PIF_definition}. Pentagons: \eq \ref{eq:n1_correct}. Filled circles: \eq \ref{eq:e_n1_final}. The PDE results were obtained using a CN timestepping scheme. $M=10^{6}$ MC realizations. Timestep $h =10^{-3}$. Parameter values: $D=0.1,~\tau_{a} = 5.0,~\widetilde{\Delta}=10,~I_{0} = 5.5, ~s_{0}^{(0)	} = 5.0.$}
\label{fig:SCC_MC_PIF}
\end{figure}

\newpage
\section{Summary and conclusions} 
In this paper, we have developed a numerical method for the computation of moments and correlations in general two-dimensional non-renewal escape time processes. Our approach relies on the numerical solution of a two-dimensional time-dependent FPE with initial conditions obtained from marginal distributions of previous states of the system. Crucially, the computation scheme presented in this study is general insofar as it can be applied to any stochastic process with a known reset condition (\eq \ref{eq:X_definition}) and any deterministic signal (\eq \ref{eq:s_definition}).
As an important application, we have described the transition to stationarity in a stochastic IF neuron model with spike-triggered adaptation, which causes non-trivial ISI correlations. A different mechanism for introducing positive correlations between ISIs has recently been reported in \cite{donofrio_pirozzi_magnasco_2016} and can equally well be analyzed with the presented methodology. Moreover, our approach enables us to determine the non-trivial timescale of transition to a stationary adapted state by counting the number of intervals needed for this transition.

Experimentally, the transition to stationarity is often characterized by the behaviour of the instantaneous firing rate \cite{la_camera_et_al_2006, naud_gerstner_2012} \footnote{Note that in \cite{la_camera_et_al_2006}, the timescale for single exponential adaptation was \textit{inferred} from the time course of the numerically obtained instantaneous firing rate, showing that the time course of the rate is not described by the same single exponential time course of the adaptation current. A similar observation is made in \cite{benda_herz_2003}.}. The instantaneous firing rate is usually obtained by averaging the neuronal activity bin-wise for a fixed time. This differs from the firing rate used here as given by the inverse of the mean ISI (\eq \ref{eq:kth_firing_time}). In other words, while the instantaneous firing rate is measured in real time, our firing rate relates to interval numbers. This entails that for a given time $t$, the firing rate contains contributions from, in general, past firing events that may have occurred at any point $k$ in the spike train. Knowing the joint distributions of all ISIs $T_{k}$, it is at least in principle possible to reconstruct the instantaneous firing rate, whereas given the instantaneous firing rate, we cannot reconstruct the joint distributions of the individual ISIs $T_{k}$. Despite the difference in the definition of the firing rate, it might be an interesting topic for further study to classify the time scales of the transition to stationarity both experimentally and based on the theory presented here.

The computation of ISI moments using the iFPT approach is computationally inexpensive, giving rise to small relative disagreements between solutions of the FPE and direct MC simulations. In contrast, the computation of correlations is harder. We observed that we lost one order of magnitude in accuracy compared to the simulation of the moments for the quantities $\mathcal Q_{1}$ (\eq \ref{eq:Q_1}) and $\mathcal Q_{2}$ (\eq \ref{eq:Q_2}), which makes the reliable computation of SCCs a computationally challenging task. We conclude that even a relative disagreement of ISI moments between Monte Carlo estimations and PDE solutions of the order $10^{-3}$ is not enough to reliably estimate the SCC using PDE numerics only (but this might be specific for the examples we have considered), indicating that more refined numerical methods or larger computational ressources, or indeed both, are needed. When the difference between the two terms in the numerator of \eq \ref{eq:SCC_definition_general} is large, the small error made by the numerical solution of the PDE should have a less detrimental influence on the final result. The need for more refined numerical methods is further substantiated by the fact that the more accurate asymptotically stable CN timestepping scheme did not result in a significant decrease in the relative disagreement between PDE results and both plain MC and MC-GS simulations, for both moments of firing intervals and the SCC. In this paper, we have only discussed the error associated with MC simulations, because it is readily available. The numerical solution of the FPE is of course also subject to numerical errors and future work will likely benefit from a discussion about how to systematically reduce these errors. In this context, it might be beneficial to compare the finite-element methods used here to other methods for solving PDEs, such as finite difference and finite volume methods \cite{rosenbaum_2012}. A systematic error estimation study might be made more difficult by the fact that the diffusion matrix (\eq \ref{eq:diffusion_matrix}) is not positive definite \cite{salsa_pde_book, ern_guermond_fem_book}.

There is an alternative method to compute the ISI distributions given the distributions of the peak adaptation currents using the formula

\begin{equation}
 \mathcal F_{k}(t) = \int_{\mathrm{supp}\left(\mathcal G_{k-1} \right)} \mathcal H(t,y) \mathcal G_{k-1}(y) \d y \, .
 \label{eq:RFPT}
\end{equation}

This is an integral equation frequently used in the context of randomized FPT problems \cite{jackson_2009, jaimungal_kreinin_valov_2014}, where usually $\mathcal F_{k}$ and the kernel $\mathcal H$ are given, and one tries to find a matching distribution $\mathcal G_{k-1}$ of starting points. Using \eq \ref{eq:RFPT}, we do not have to solve a time-dependent PDE for each ISI, but must compute $\mathcal H$ once as the solution of a time-dependent FPE with varying initial conditions for $s$, similar to the computation of $\mathcal F_{1}$. The averaging that the iFPT approach amounts to is particularly clear in this formulation. The densities $\mathcal G_{k}$ are obtained as discussed above (see \eq \ref{eq:density_z_integration}). The approach relying on \eq \ref{eq:RFPT} might be computationally less expensive, but we found that it is not as exact as solving a time-dependent FPE for each ISI, especially at larger times. This is likely caused by errors when computing $\mathcal H$, as the numerical integration in \eq \ref{eq:RFPT} can be performed accurately and efficiently. However, \eq \ref{eq:RFPT} could be useful for analytical explorations when $\mathcal H$ is known.

We finally emphasize that our approach did not use the complicated boundary conditions for stationary IF models, where the probability \textit{flux} at threshold gives rise to a discontinuity of the probability flux at reset \cite{brunel_sergi_1999, schwalger_linder_pre_2015}. In contrast, our approach allows for the computation of transient and stationary distributions of the adaptation dynamics in an iterative fashion, requiring the solution of a two-dimensional time-dependent PDE. The only boundary condition that has to be taken into account is an absorbing boundary condition for the probability \textit{density} at the threshold $\xth$. This makes the problem tractable using finite-element approximation methods for time-dependent PDEs, resulting in a general description of two-dimensional IF models with spike-triggered adaptation. The approach we have described in this paper can in principle also be used to gain analytical insight into these system, however, quantities such as $\mathcal H$ and the solution of a two-dimensional time-dependent PDE seem to be unavailable in closed analytical form except in the most simple cases.

\begin{acknowledgments}
We would like to thank Alexandre Payeur for insightful discussions and helpful comments (especially about the equivalence of \eqs \ref{eq:n1_correct} and \ref{eq:n1_traditional}) and NSERC Canada for funding this work. Furthermore, we would like to thank an anonymous referee for comments that helped us to improve the manuscript.
\end{acknowledgments}

\appendix
\section{Equivalence of \eqs \ref{eq:n1_correct} and \ref{eq:n1_traditional}}
\label{sec:appendix}
We here show that \eq \ref{eq:n1_correct} and \eq \ref{eq:n1_traditional} are equivalent.

We recall \eq \ref{eq:n1_correct}:
\begin{align}
\label{eq:n1_correct_2}
&\mathbb{E}(T_nT_{n+1})=\\
&\int \d T_n\,\d T_{n+1}\,\d s_{0}^{(n)}\,T_n\,T_{n+1}\,\mathcal{H}(T_{n+1},s_{0}^{(n)})p(s_{0}^{(n)},T_n) \, . \nonumber
\end{align}

We re-write  \eq \ref{eq:n1_traditional} as follows:
\begin{equation}
\begin{split}
&\mathbb E \left( T_{n} T_{n+1}\right) = \\
&\int\d T_{n} \,\d T_{n+1}\,\d y \, T_{n}\,T_{n+1}\,\mathcal H\left(T_{n+1}, f(T_{n}, y)\right) p(y, T_{n})  \, ,
\end{split}
\label{eq:n1_traditional_2}
\end{equation}

where $y=s_{0}^{(n-1)}$ and we have replaced $\mathcal H(T_{n}, y) \mathcal G_{n-1}(y) = p(y, T_{n}) $. Note that in \eq \ref{eq:n1_correct_2}, $p$ is the joint density of $s_{0}^{(n)}$ and $T_n$, whereas $p$ is the joint density of $s_{0}^{(n-1)}$ and $T_n$ in \eq \ref{eq:n1_traditional_2}.

By inspection, the two expressions are identical if we can show that

\begin{equation}
\begin{split}
&\int \d s_{0}^{(n)} \mathcal H(T_{n+1}, s_{0}^{(n)}) p(s_{0}^{(n)}, T_{n}) = \\
&\int \d s_{0}^{(n-1)} \mathcal H(T_{n+1}, f(T_{n}, s_{0}^{(n-1)}))p(s_{0}^{(n-1)}, T_{n}) \, ,
\end{split}
\label{eq:app_1}
\end{equation}

for $T_{n}$ and $T_{n+1}$ fixed.

Starting from the second line in \eq \ref{eq:app_1}, we change the integration variable from $s_{0}^{(n-1)}$ to $s_{0}^{(n)}$ by observing that from $s_{0}^{(n)} = f(T_{n}, s_{0}^{(n-1)})$, we have $\frac{\d s_{0}^{(n)}}{\d s_{0}^{(n-1)}} = \frac{\partial f}{\partial s_{0}^{(n-1)}}$ and therefore $\d s_{0}^{(n-1)} = \d s_{0}^{(n)} \left(\frac{\partial f}{\partial s_{0}^{(n-1)}}\right)^{-1}$. We need to assume that $f$ is invertible with respect to the second argument, which is the case for both power-law (\eq \ref{eq:f_definition}) and exponential adaptation (\eq \ref{eq:f_definition_exponential}) considered in this paper. The integral then becomes

\begin{widetext}
\begin{equation}
\int\d T_{n} \,\d T_{n+1}\,\d s_{0}^{(n)} \, T_{n}\,T_{n+1}\,\mathcal H\left(T_{n+1}, s_{0}^{(n)})\right) p\left[f^{-1}(T_{n}, s_{0}^{(n)}), T_{n}\right] \left(\frac{\partial f}{\partial s_{0}^{(n-1)}}\right)^{-1} \, .
\end{equation}
\end{widetext}

But $ p\left[f^{-1}(T_{n}, s_{0}^{(n)}), T_{n})\right] \left(\frac{\partial f}{\partial s_{0}^{(n-1)}}\right)^{-1} $ is nothing but the transformation from $p(s_{0}^{(n-1)}, T_{n})$ to $p(s_{0}^{(n)}, T_{n})$. Indeed, we have (fixing $T_{n}$)

\begin{equation}
 p(s_{0}^{(n)}, T_{n}) \frac{\partial s_{0}^{(n)}}{\partial s_{0}^{(n-1)}}=  p(s_{0}^{(n-1)}, T_{n}) \, ,
\end{equation}

so that

\begin{equation}
  p(s_{0}^{(n)}, T_{n})  = p\left[f^{-1}(T_{n}, s_{0}^{(n)}), T_{n}\right]\left(\frac{\partial f}{\partial s_{0}^{(n-1)}}\right)^{-1} \, .
\end{equation}

Therefore, \eq \ref{eq:n1_correct} and \eq \ref{eq:n1_traditional} are equivalent.

\bibliography{literature_paper}

\end{document}